\documentclass[a4paper,11pt]{article}
\usepackage{jinstpub} 
\usepackage{lineno}

\usepackage{lineno}
\usepackage{subcaption}
\usepackage{enumitem}
\usepackage{soul}

\newcommand{\commentOut}[1]{\iffalse #1 \fi}

\title{Neutron and \texorpdfstring{$\boldsymbol{\gamma}$}{}-ray Discrimination by a Pressurized Helium-4 Based Scintillation Detector}

\author[1]{Shubham Dutta}
\emailAdd{shubhamdutta\_16@yahoo.com}
\author[2]{Sayan Ghosh}
\author[3]{Satyajit Saha}

\affiliation[1]
{   High Energy Nuclear and Particle Physics Division,
    Saha Institute of Nuclear Physics - a CI of Homi Bhabha National Institute,
    Block AF, Sector I, 
    Bidhannagar, Kolkata,
    PIN: 700064, 
    West Bengal, India
}

\affiliation[2]
{   Kavli Institute for Particle Astrophysics and Cosmology,
    Stanford University,
    Stanford, CA 94305, USA
}

\affiliation[3]
{   Applied Nuclear Physics Division,              
    Saha Institute of Nuclear Physics - a CI of Homi Bhabha National Institute,
    Block AF, Sector I, 
    Bidhannagar, Kolkata,
    PIN: 700064, 
    West Bengal,
    India
}

\abstract{
Pressurized helium-4 (PHe) based fast neutron scintillation detector offers an useful alternative to organic liquid-based scintillator due to its relatively low response to the $\gamma$-rays compared to the latter type of scintillator. In the present work, we have investigated the capabilities of a PHe detector for the detection of fast neutrons in a mixed radiation field where both the neutrons and the $\gamma$-rays are present. Discrimination between neutrons and $\gamma$-rays is achieved by using fast-slow charge integration method. We have also conducted systematic studies of the attenuation of fast neutrons and $\gamma$-rays by high-density polyethylene (HDPE). Additionally, the simulation analyses, conducted using GEANT4, provide detailed insights into the interactions of the radiation quanta with the PHe detector.
}


\begin{document}
\maketitle
\flushbottom

\section{Introduction}
\label{Intro}
Neutrons as radiation quanta, are found in nature because of their release in various forms of nuclear reactions, most common being the nuclear fission and the $(\alpha,n)$ reactions caused by the natural radioactivity of the remnants of Uranium-Thorium (U-Th) decay chain. Free neutrons are, however, unstable against $\beta$-decay with about 15 minutes of half-life. In spite of this fact, free neutrons pass and penetrate through matter since they are charge neutral and deposits partial or full energy through hadronic interactions. These interactions include capture into nuclei and elastic scattering, resulting in nuclear recoil of the stopping media. Such interactions take place over a very short time scale compared to the half-life of free neutrons, which make it possible to detect them in real life.

Neutrons are produced in large number inside the core of the nuclear reactors and also emerge out of the spent nuclear fuels by spontaneous fission and $(\alpha,n)$ reactions\cite{Runk}. These neutrons, detected with the help of suitable neutron detectors, are often used to monitor the spent fuel repositories and also at the strategic surveillance stations for monitoring hidden nuclear materials. In that respect, neutron detectors, capable of detecting neutrons up to around 10 MeV energy, serve the purpose. However, since the neutrons are most often found in a mixed radiation field with dominance of mostly $\gamma$-rays, X-rays and electrons coming out of the same sources, it is important to achieve discrimination between the neutrons and the other radiation quanta before any meaningful result can be extracted.

Neutron background for a typical dark matter search experiment, usually set up at underground laboratories, interferes with the signal due to possible dark matter candidates, as both the radiation quanta interact with active media to produce overlapping signals. These neutrons are predominantly produced by the $(\alpha,n)$ reactions due to the U--Th decay chain products emitting $\alpha$-particles. Careful measurements of the neutron background is essential at every site to assess the sensitivity limits. Pressurized helium-4 (PHe) detector has been used recently at such facilities to monitor the residual neutron flux\cite{Sayan}.

Liquid Helium\cite{Thor} has been investigated as scintillator for neutron detection more than 60 years back. However, the scintillation properties of PHe gas have been examined much later for successful implementation as fast neutron detector\cite{Chan}. The major advantage of liquid Helium-4 and PHe gas as scintillators is relatively weak response to $\beta$-particles and  $\gamma$-rays, because the density of available electrons in Helium is much less than in standard scintillators, organic and inorganic. On the other hand, neutrons cause nuclear recoil of the Helium-4 nuclei inside the pressurized gas, resulting in multiple processes of ionization and other interactions, leading to scintillation through transitions from the singlet excimer states or the triplet excimer states\cite{Kell}. Both the transitions lead to emission in the extreme ultrared (EUV) region of the electromagnetic spectrum, with wavelength of maximum emission around 80 nm. A wavelength shifting (WLS) compound is used to make the EUV scintillation light output readable by photomultipliers (PMTs) or silicon photomultipliers (SiPMs)\cite{Mure,Jeba}.  

\begin{figure}[htbp!]
    \centering 
    \vspace{-1.\baselineskip}
    \includegraphics[height=75mm]{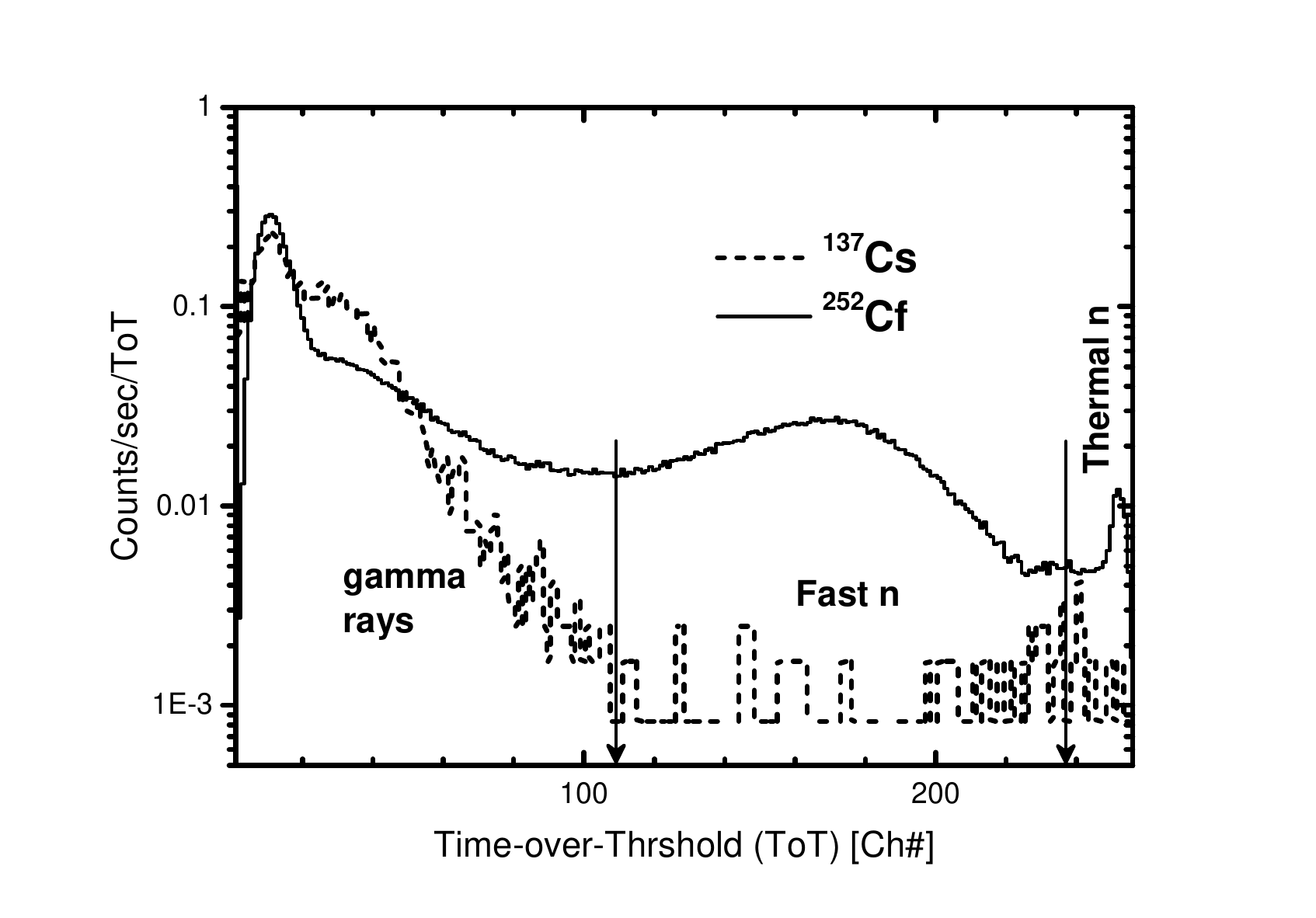} 
	\caption{Time over threshold (ToT) plots obtained by exposing the detector separately with a $^{252}$Cf gamma ray and neutron source, and a $^{137}$Cs gamma ray source. The two arrows mark the low ToT and the high ToT cuts, which separate the regions of response for the gamma rays, the fast neutrons and the thermal neutrons as shown.}
	\label{fig00}
 \end{figure}

In recent times, PHe detectors are made commercially available by Arktis Radiation Detectors Ltd., Switzerland\cite{arktis}. Front-end electronics and data acquisition system is provided as a package for measuring the fast and the thermal neutron fluxes after pulse shape discrimination using Time-over-Threshold (ToT) technique to discriminate between $\gamma$-rays, fast neutrons and the thermal neutrons. The inner wall of the detector is coated with a Lithium-based compound to make the detector sensitive to the thermal neutrons\cite{arktis}. These neutrons undergo capture by Lithium to produce energetic $\alpha$-particles which result in scintillation inside the detector volume. Corresponding ToT signals are found to be larger than those produced by the fast neutrons. ToT spectra measured with the S670e PHe detector for the neutrons and the $\gamma$-rays emitted from a $^{252}$Cf neutron source and a $^{137}$Cs $\gamma$-ray source are shown in the Fig.~\ref{fig00}. Two different cuts along the ToT axis, shown by the arrows, differentiate between the $\gamma$-rays, fast neutrons and the thermal neutrons as labelled in the figure. One disadvantage of this method of discrimination is considerable overlap between the gamma rays and the fast neutrons or between the fast neutrons and the thermal neutrons. As a result, a clear threshold for discrimination could not be established.

Intrinsic detection efficiency of the above-mentioned PHe detector for the neutrons has been extensively studied using neutron Time-of-Flight (ToF) experiment\cite{LIANG20191} over the energy range of $\sim0.5 - 7$~MeV\cite{LIANG20191}. Maximum efficiency was found to be 6.8\% around 
1.5~MeV. It varies within the range of $\sim4 - 7\%$ over the energy range of 1 - 7~MeV. However, the simulated efficiency values over the same range was found to differ widely, peaking around 11\% at $\sim 1$~MeV.

The objectives of the present work are:
\vspace{-1.0\baselineskip}
\begin{itemize}[noitemsep]
    \item discrimination of the fast neutrons from the $\gamma$-rays and electrons using fast-slow charge integration method;
    \item estimation of threshold for neutron and $\gamma$-ray discrimination;
    \item qualitative study of attenuation of the fast neutrons and $\gamma$-rays by high density polyethylene (HDPE or PE);
    \item carry out a detailed simulation to a) establish the neutron emission spectrum from a $^{241}$Am based radioactive source used in the experiment, b) to understand the interaction of the neutrons and $\gamma$-rays with the PHe media;
    \item follow the systematics of energy transfer from the neutrons and the $\gamma$-rays to $^4$He (nuclear recoil) and electrons (electron recoil) respectively. 
\end{itemize}

\section{Experimental Details}
\label{expt}

\subsection{Instrumentation}
\label{Instr}
\begin{figure}[ht]
    \begin{subfigure}{0.5\textwidth}
	    \centering 
	    \includegraphics[scale=0.25]{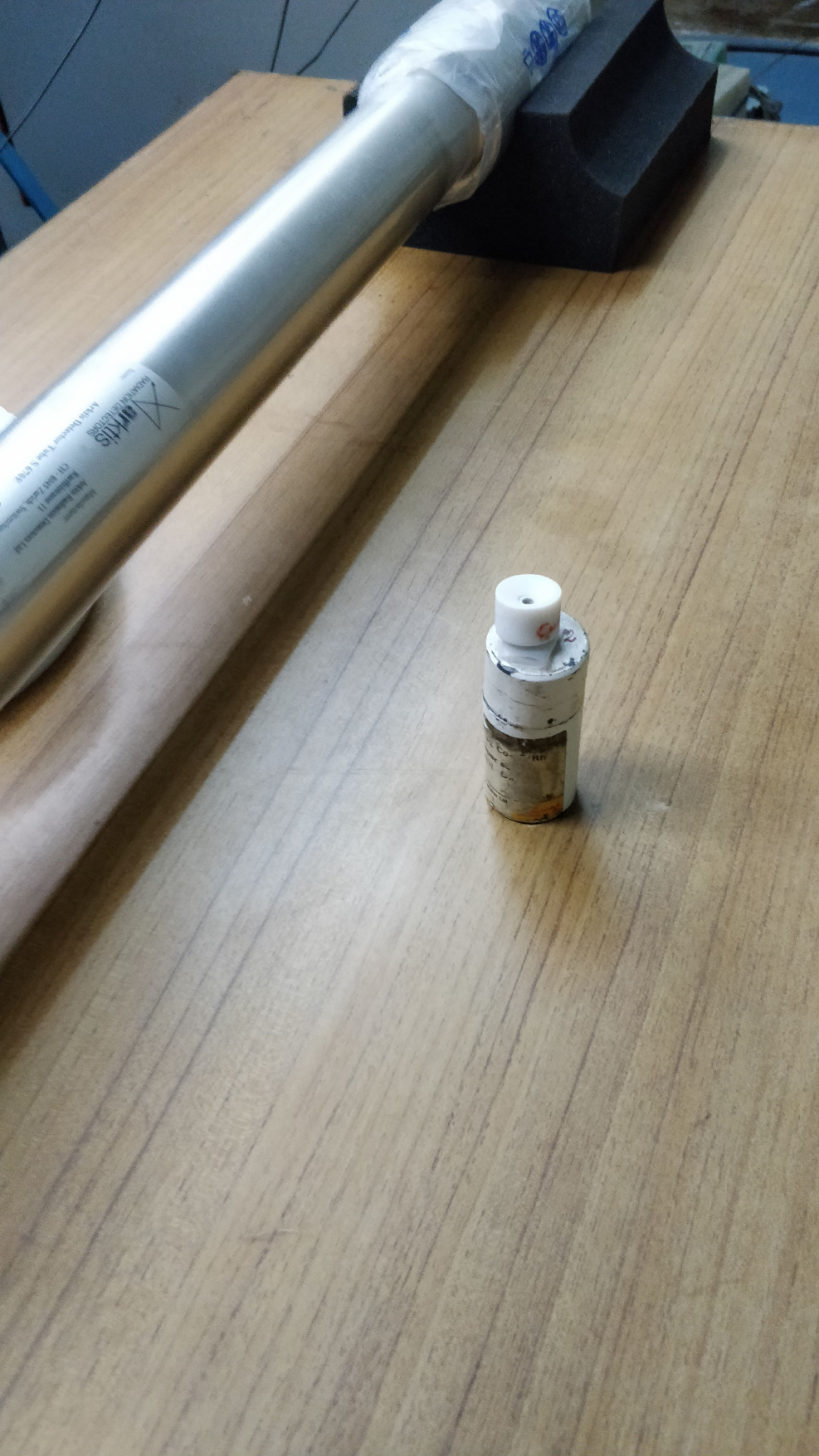}
	    \caption{}
        \label{fig1a}  
    \end{subfigure}%
    \begin{subfigure}{0.5\textwidth}
	      \centering  
        \includegraphics[scale=0.25]{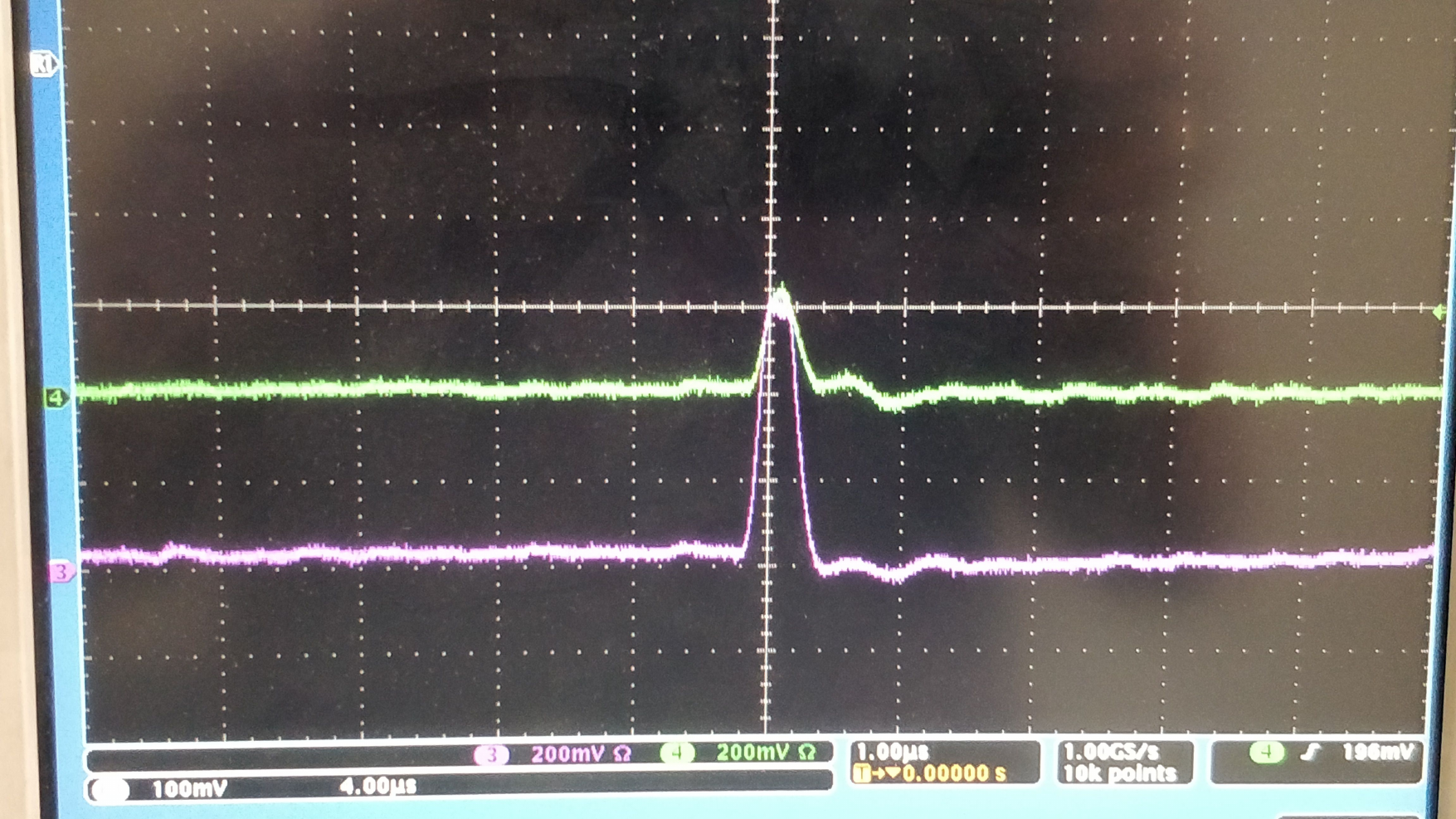}
        \caption{}
        \label{fig1b}
    \end{subfigure}
	\caption{(a) Experimental set up involving Arktis detector and $^{252}$Cf source inside teflon capsule; (b) signal traces of the amplified pulses from the detector.}
	\label{fig1}
\end{figure}

A photograph of the experimental set up is shown in the Fig.~\ref{fig1a}. Arktis-made PHe detector and a radioactive source ($^{252}$Cf source inside a PTFE capsule), placed in front of the detector, are displayed in the photograph. The active medium of the detector is $^{4}$He gas at a pressure of 180 bar, enclosed inside a stainless steel tube of 60 mm inner diameter and 600 mm active length. The detector is packaged with SiPM arrays as photon readout located inside the pressurized detector volume. The detector is segmented into three parts along the length of the cylindrical tube. Photons from each segment are read out by an array of 8 SiPMs. Signals from two SiPMs are summed and fed into each output, so that 4 output signal pulses are generated from each segment. Details about the detector, related electronics and their working principles are given in many references available\cite{Mure,arktis}.

For our work, the digital electronics readout system was replaced by a multi-channel analog circuit (provided by the manufacturer) for analog signal processing using conventional electronics and a 12-bit current-integrating analog-to-digital converter or QDC (16 channel Phillips Scientific charge ADC, Model 7187). The pulses from the SiPMs are fed to the analog circuit, which consists of a preamplifier and a shaping amplifier with baseline restorer for each SiPM channel. Traces of the output pulses from the amplifier are shown in the Fig.~\ref{fig1b}. The output signals were inverted using high speed pulse transformer (ALT 4532M) to match with the polarity and pulse timing requirements. Out of the four output signals from a segment, one signal was fed to a low threshold discriminator to generate the logic gates for the QDCs required for pulse shape discrimination (PSD) between neutrons and $\gamma$-rays or electrons. The logic gates, {\em viz.} the long gate was matched with the timing and the duration of the output signal in order to integrate the whole pulse, while the short gate was chosen by optimization of the PSD signal. The PSD parameter ($P$) is defined as: 
$P=N Q_S/{(Q_S+Q_L)}$ where, $Q_S$ and $Q_L$ are the charge contents of the pulse within the duration of the short gate and the long gate respectively and $N$ is a scaling factor to convert the ratios to suitable integer values for the plots.   

\subsection{Measurements with Radioactive sources}
The systematic studies were done using different neutron and $\gamma$-ray sources placed at a certain distance from the detector as shown in the photograph (see Fig.~\ref{fig1a}). Provision for placement of different absorber materials in between for systematic studies was also made. An unmoderated $^{252}$Cf spontaneous fission source (half-life = 2.645 years, spontaneous fission branching ratio = 3.09\%\cite{wats}) was used for simultaneous detection of neutrons and $\gamma$-rays. The source, with a residual strength of a few kBq, emits fast neutrons with average energy of 2.12 MeV. The spontaneous fission source also emits $\gamma$-rays of various energies ranging within a broad range of $0.1 - 1$~MeV due to the transitions between the nuclear levels of the residual fission fragments. Since the source had already passed through approximately 8 half-lives since its preparation, the emission rates of $\gamma$-rays compared to that of fast neutrons are higher by at least 2 orders of magnitude.   
The source  was sealed inside a cylindrical PTFE capsule having an opening on one of the flat faces. The aperture was sealed with a 100 $\mu$m thick polyethylene terepthalate (PET) window to prevent the alpha particles, recoiling heavy nuclei including the fission fragments from ejection into the air.

\begin{figure}[htbp!]
\begin{subfigure}{0.5\textwidth}	
 \centering
	\includegraphics[height=60mm]{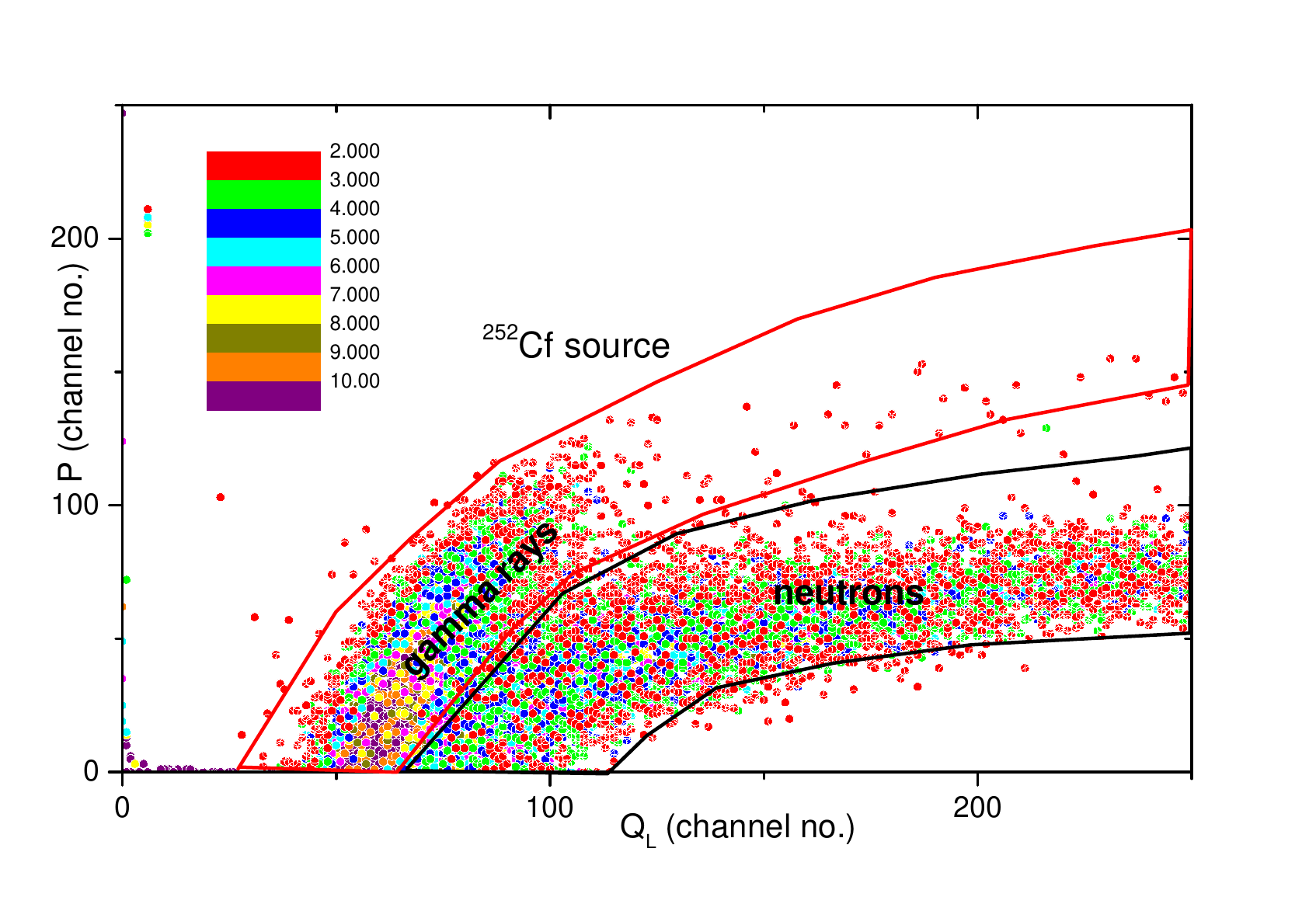}
 \end{subfigure}
 \begin{subfigure}{0.5\textwidth}	
     \includegraphics[height=60mm]{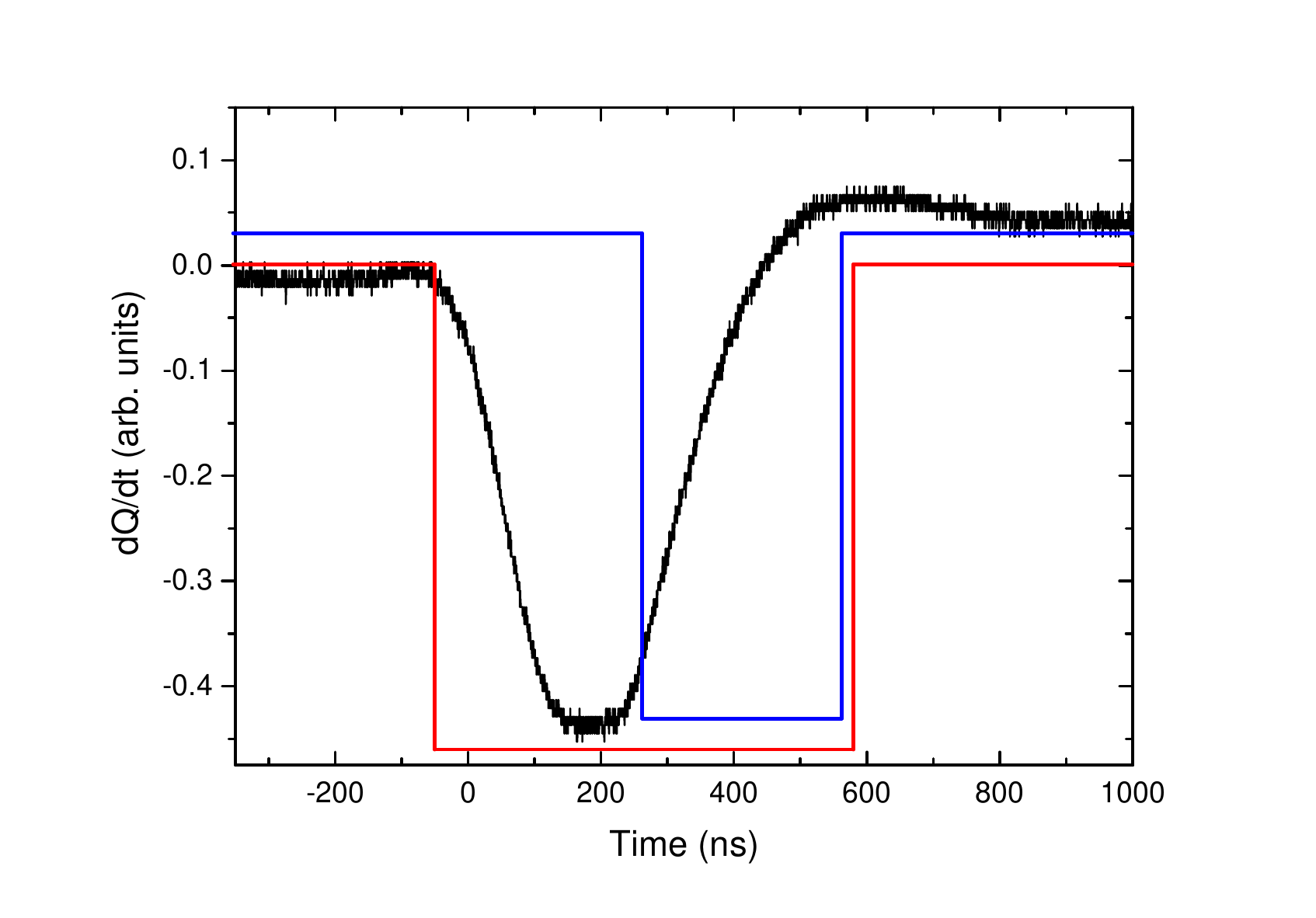}
 \end{subfigure}
\caption{(a) 2D scatter plot of the PSD parameter $P$ as function of $Q_L$ obtained by exposure of $^{252}$Cf source. The contours marked around the bands are the banana gates for the neutrons (black) and the $\gamma$-rays (red). (b) A typical signal pulse output of the detector with the short gate (blue) and the long gate (red) is shown.}
\label{fig2}
\end{figure}

A typical two dimensional (2D) isometric or scatter plot of the PSD parameter $P$ vs. $Q_L$ is shown in the Fig.~\ref{fig2} for the $^{252}$Cf source placed near the detector. Two distinct bands can be seen in the scatter plot 
obtained after 5 hours of exposure to the $^{252}$Cf source. The relative configuration of the gates, optimized by adjusting gate widths and delays by looking at $P$, is also shown in Fig.~\ref{fig2}b. Optimum delay between the threshold for the up-swing of the pulse and the trigger point of the short gate was found to be around 300 nanosecond for achieving good discrimination. 

\begin{figure}[ht]
	\centering
    \vspace{-1\baselineskip}
	\includegraphics[height=60mm]{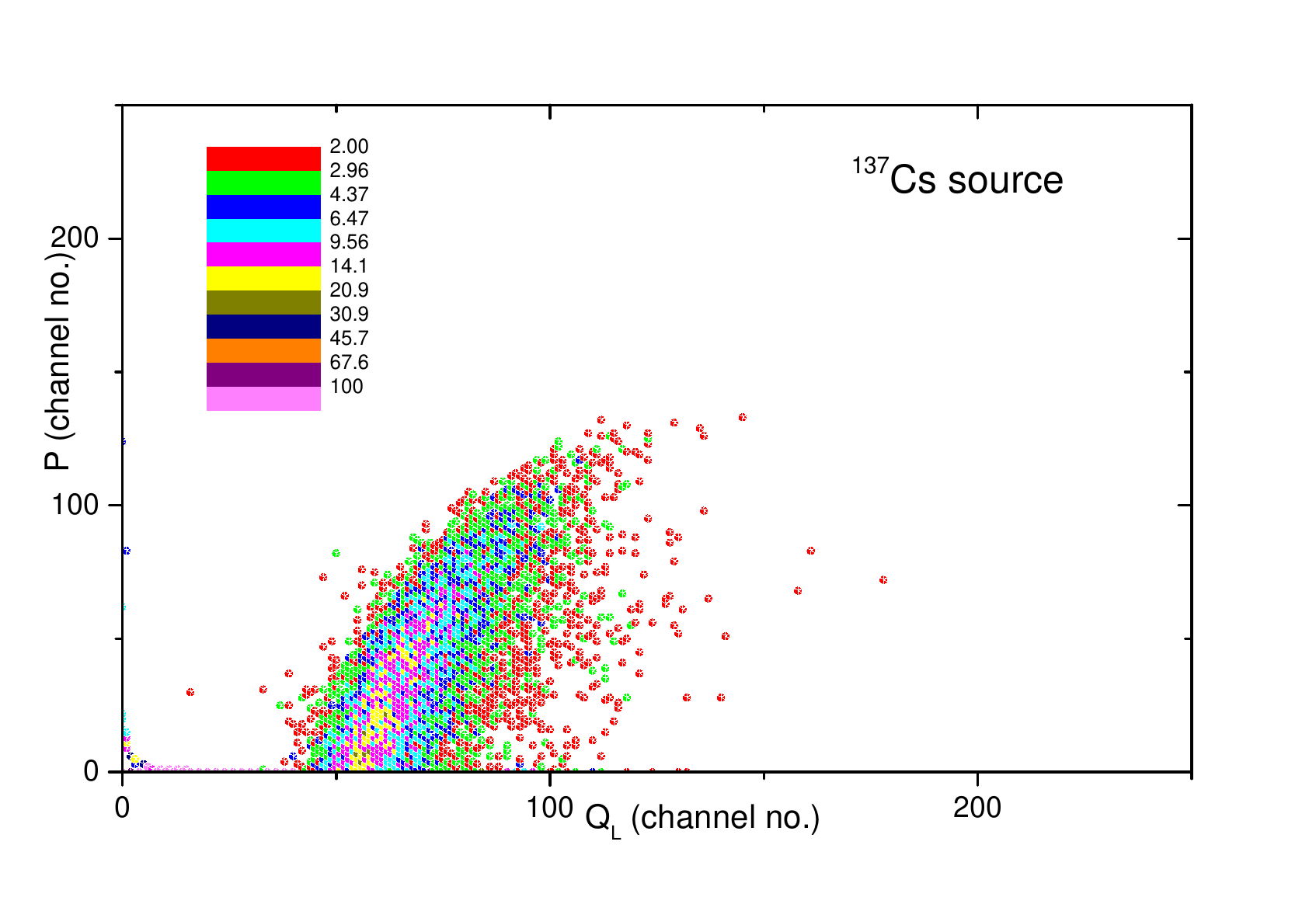}
	\caption{Scatter plot of the PSD parameter $P$ as function of $Q_L$ obtained by exposure of a $^{137}$Cs $\gamma$-reference source.}
    \label{fig3}
\end{figure}

A $^{137}$Cs monoenergetic (662 keV) $\gamma$-ray source, with strength $\sim30\,{\rm kBq}$, was used for the experiment to identify the $\gamma$-band as distinct from the neutron band. The scatter plot, obtained after exposing the detector with the $\gamma$-ray source, is displayed in Fig.~\ref{fig3}, which shows a single band ($\gamma$-band) as expected. Additional confirmation of the neutron band was obtained after systematic attenuation studies made using high density polyethylene (HDPE or PE) absorbers (see Sec.~\ref{HDPEstudies}).

\begin{figure}[htbp!]
	\centering
    \includegraphics[height=90mm]{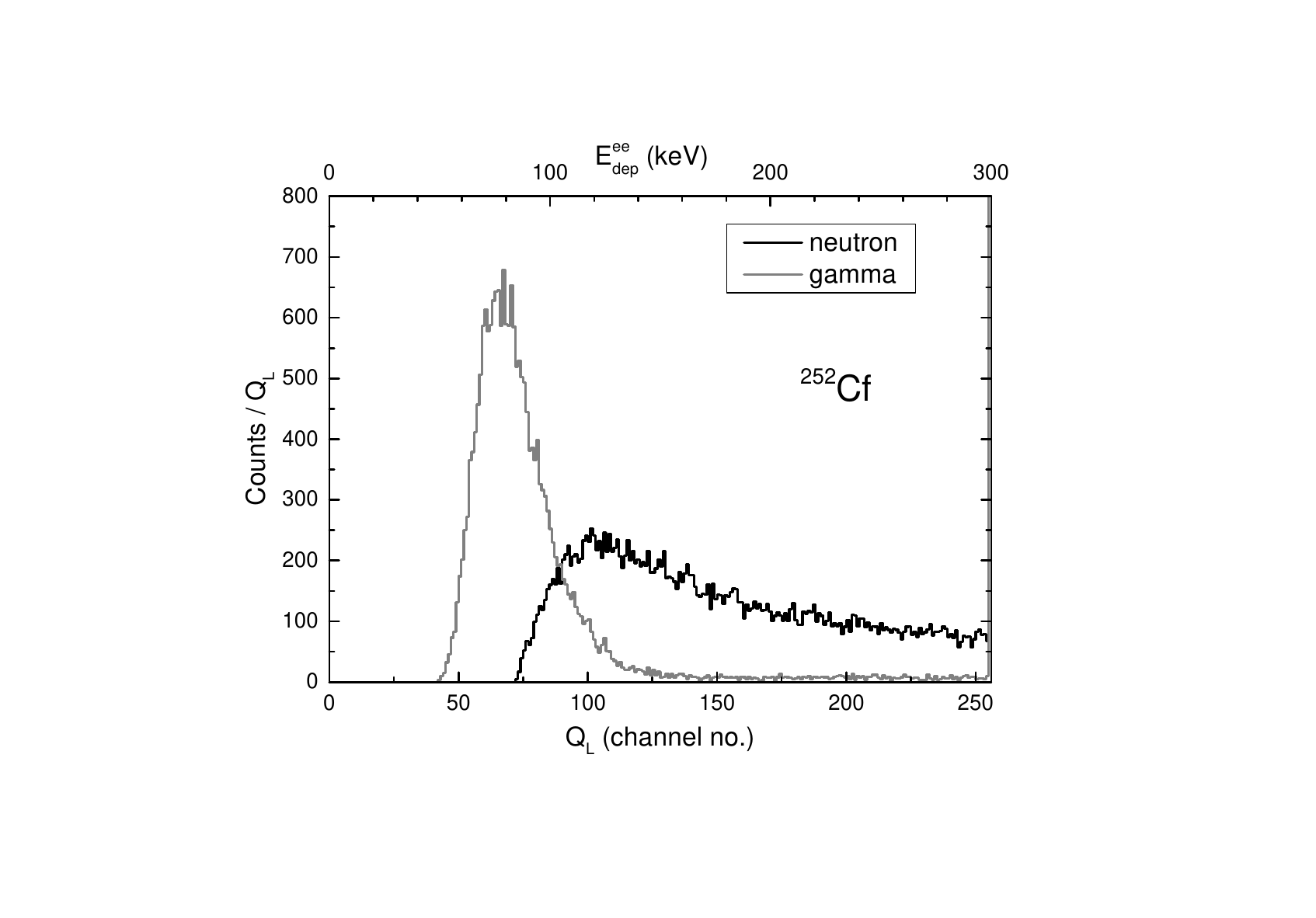}
    \vspace{-2.5\baselineskip}
	\caption{Projection spectra of the neutron band and the $\gamma$-band obtained by gating on the respective bands for exposure to the $^{252}$Cf source.}
	\label{fig4}
\end{figure}

From the 2D scatter plot of Fig.~\ref{fig2}a, the neutron and the $\gamma$-bands are found to merge at low $Q_L$ values, which qualitatively indicates the low energy threshold for discrimination. Projected spectra of the neutrons and the $\gamma$ bands were taken after selection of the corresponding bands by the contours (popularly termed as banana gates) shown in the 2D scatter plots of Fig.~\ref{fig2}a. A comparison of projection of the neutron and the $\gamma$-bands on the $Q_L$ axis is shown in Fig.~\ref{fig4}. It reveals that the detector response to the $\gamma$-rays, mediated mostly through electron recoil, is quite low compared to that for the neutrons having energies of the same order. It can be seen from the plots that the nuclear recoil spectrum due to the neutrons terminate abruptly at higher channels ($\gtrsim 250$), which is due to the saturation of the pulses of the SiPM signal amplifiers provided by the manufacturer. 

\subsection{Measurements with \texorpdfstring{$^{241}$}{}Am source}
\label{241Am}
A $^{241}$Am source (432 years half-life), embedded in a glass matrix and sealed inside a cylindrical stainless steel capsule, was also studied with the detector. The source, having a strength of $\sim100 \, {\rm MBq}$, is originally intended for energy calibration of X-ray detectors. This source was used in the experiment to determine the response of the detector to low energy photons so that a low energy cut-off for electron recoil could be established. The scatter plot of $P$ vs. $Q_L$, obtained after 5 hours of exposure, is shown in Fig.~\ref{fig6a}. A clear neutron band, besides the $\gamma$-band, was observed to appear. This was a bit surprising due to the reason that 1) the source was used so far for energy and relative efficiency calibration of X-ray detectors, and 2) it was never monitored with neutron detectors or dosimeters. The same exposure time (5 hours) for the $^{241}$Am and the $^{252}$Cf sources to the PHe detector was used.

In order to understand the origin of the neutrons and its associated spectral distribution, a detailed simulation was carried out to understand the production of neutrons from the $^{241}$Am radioactive source. This was done using the known geometry and the media surrounding the source material as described in the following paragraph. The simulation study was performed using GEANT4 (G4) simulation toolkit \cite{geant4}, version 10.7.4. A custom physics-list is utilized in our simulation, which is based on the one developed by Mendoza et al.\cite{Mendoza2020}. We will briefly outline the G4 packages used in this physics-list to model various interactions. The G4RadioactiveDecayPhysics package is used for modeling of the radioactive decays, relevant for the $^{241}$Am source. It uses the ENSDF database \cite{Tuli1996} for the various decay parameters including the energy levels of the daughter nuclei. The production of neutrons from ($\alpha,n$) reaction is modeled using the G4ParticleHP package, which is capable of utilizing the ENDF-6 formatted data libraries\cite{ENDF6}. The database used for this purpose is the JENDL/AN-2005 dataset \cite{JENDLAN2005}. The QGSP\_BIC\_HP model is used for the hadronic interactions and uses the G4 Neutron Data Library (G4NDL) to implement low energy neutron interactions with high-precision. The EM interactions are modeled with the G4EmStandardPhysics\_option4 package.

The geometry to obtain the particle spectra consists of a cylindrical stainless steel source capsule of 6~mm diameter $\times$ 9~mm length, with the $^{241}$Am source, embedded in a pyrex glass matrix of 3~mm diameter $\times$ 6~mm thickness, placed inside and sealed with a stainless steel cover. The source capsule is surrounded by a spherical {\em dummy} detector, which is made of air. This is placed to track the particles that are being emitted from the source. $^{241}$Am is used as the primary $\alpha$-emitter nucleus embedded within the glass matrix. It decays to $^{237}$Np by emitting $\alpha$-particles at 5.486 MeV ($\sim85\%$ decay branch), 5.443 MeV ($\sim13\%$ decay branch), and the rest at other energies. Np$^{237}$ daughter nucleus has a half-life of approximately 4 orders of magnitude more than that of $^{241}$Am. Therefore, the source is approximated as an alpha emitter whose energy is sampled from the spectrum as obtained from G4. These $\alpha$-particles are allowed to penetrate isotropically through the base material of the source capsule, where they undergo ($\alpha,n$) reaction with the constituent nuclei. Based on the cross-section data available from the database, respective elemental compositions of the materials and the relevant energies of the $\alpha$-particles, we find that the dominant reaction producing neutrons are: $^{11}$B($\alpha, n$)$^{14}$N ($\sim69\%$) and $^{23}$Na($\alpha$, n)$^{26}$Al($\sim11\%$), rest are being produced by various other reactions.

\begin{figure}[ht] 
    \centering
    \vspace{-1\baselineskip}
	\includegraphics[height=55mm]{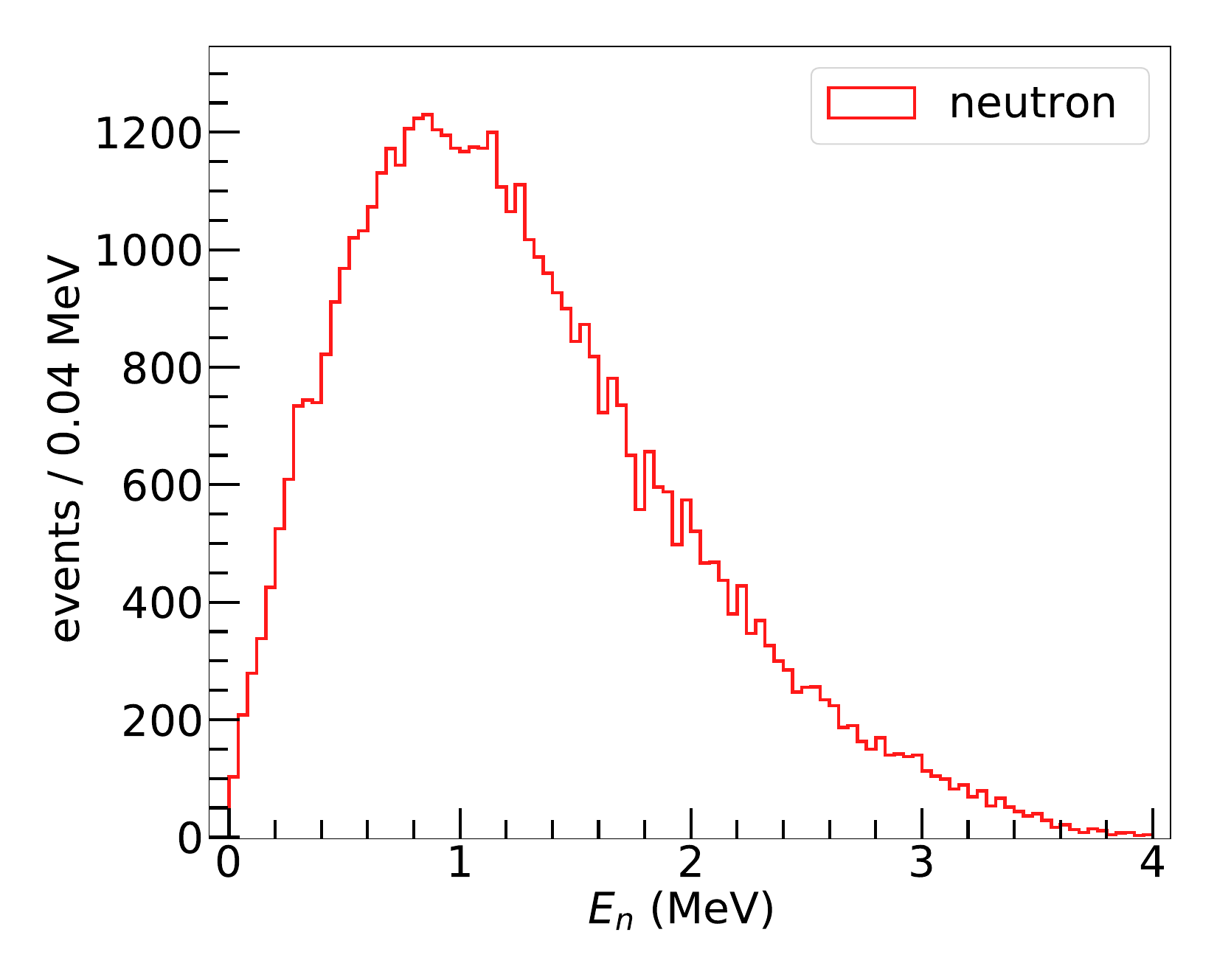}	
    \vspace{-1\baselineskip}   
	\caption{Energy spectrum of the neutrons arising from the $^{241}$Am source capsule as obtained from G4 simulation.}
	\label{fig9}
\end{figure}

The neutron yield from these ($\alpha,n$) reactions is found to be very low, $\sim1$ per million for the impinging $\alpha$-particles. Accumulating sufficient statistics would require significant amount of computation time. The physics-list has the option to bias the ($\alpha, xn$) cross-section by a fixed factor in order to increase the neutron yield. The developers of the physics-list have tested and verified that the biasing technique do not impact the neutron energy spectrum \cite{Mendoza2020}. A biasing factor of 10000 was used. The resulting neutron spectral distribution is shown in the Fig.~\ref{fig9}. The neutron spectrum has a peak at $\sim1$~MeV and extends up to $\sim4$~MeV.

\subsection{Systematic Studies with HDPE Attenuator}
\label{HDPEstudies}
\begin{figure}[htbp!]
    \begin{subfigure}{0.5\textwidth}
	    \centering
        \includegraphics[height=50mm]{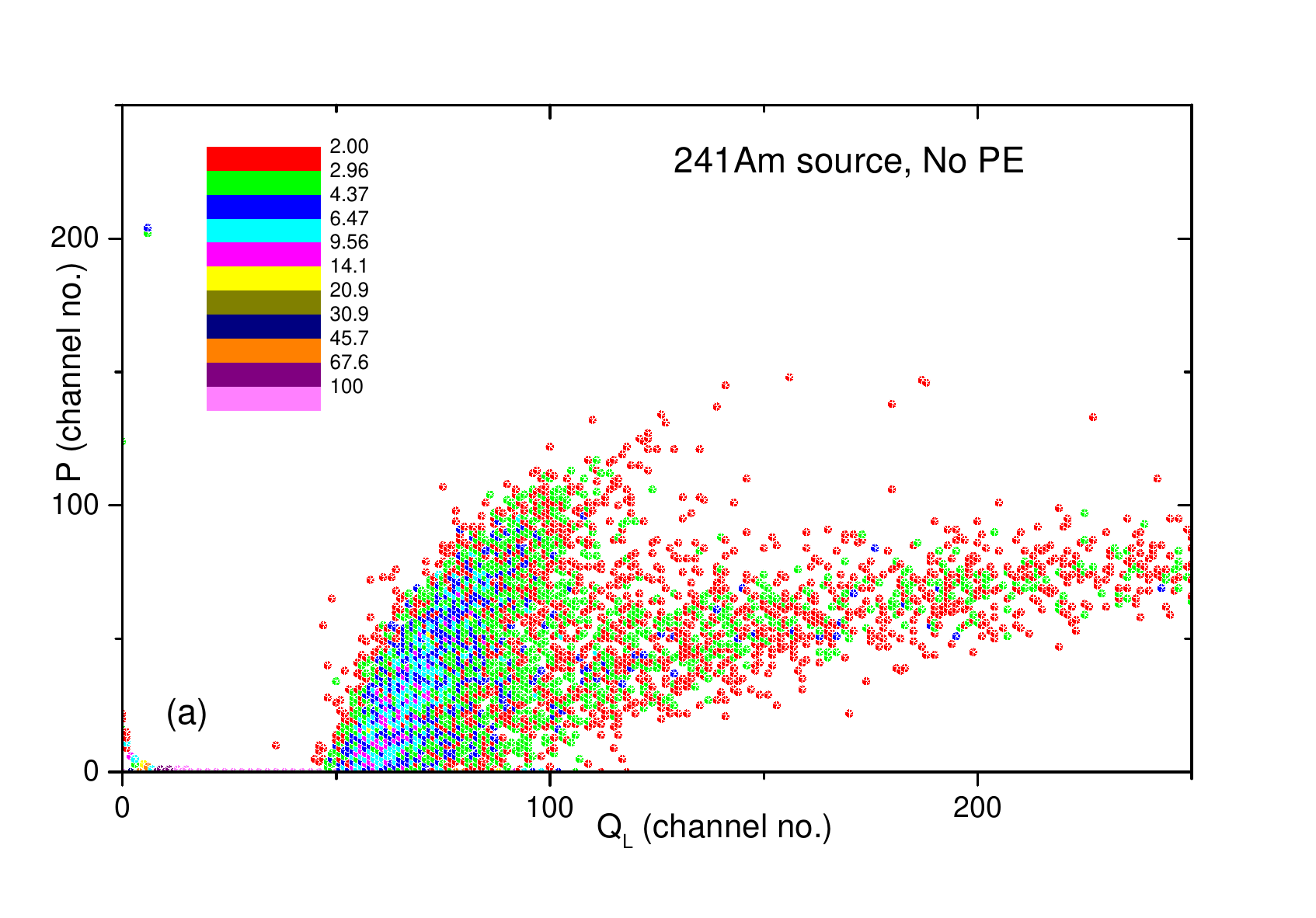}
        \phantomsubcaption
	    \label{fig6a}
    \end{subfigure}%
    \begin{subfigure}{0.5\textwidth}
        \centering
        \includegraphics[height=50mm]{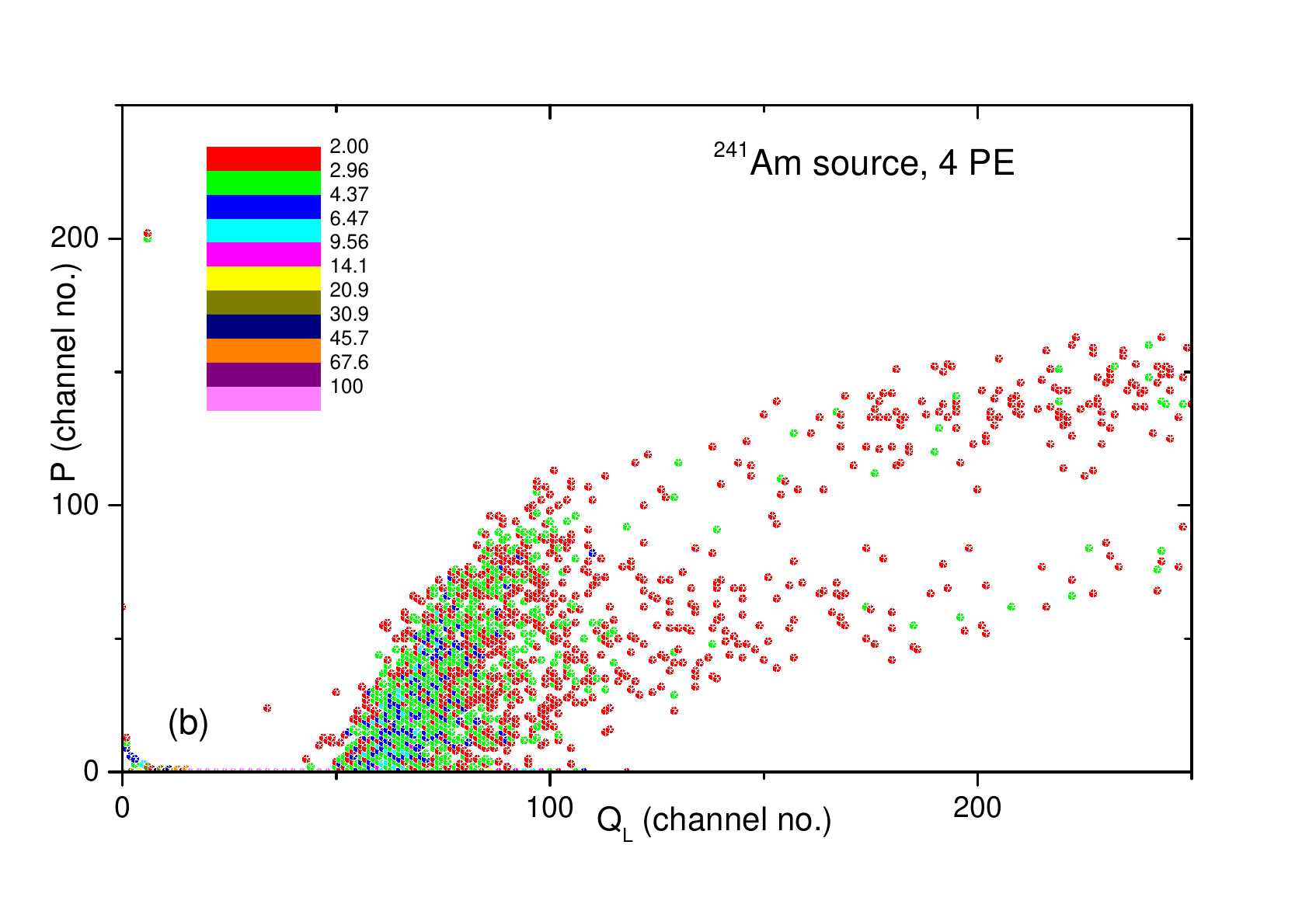}
        \phantomsubcaption
        \label{fig6b}        
    \end{subfigure}	
    \vspace{-1\baselineskip}
	\caption{(a) Scatter plot of the PSD parameter $P$ as function of $Q_L$ obtained by exposure of a sealed $^{241}$Am source; (b) same plot obtained after blocking the neutrons by 100 mm thick HDPE absorber.}
	\label{fig6}
\end{figure}

\begin{figure}[htbp!]
    \begin{subfigure}[t]{0.5\textwidth}
	    \centering
	    \includegraphics[height=55mm]{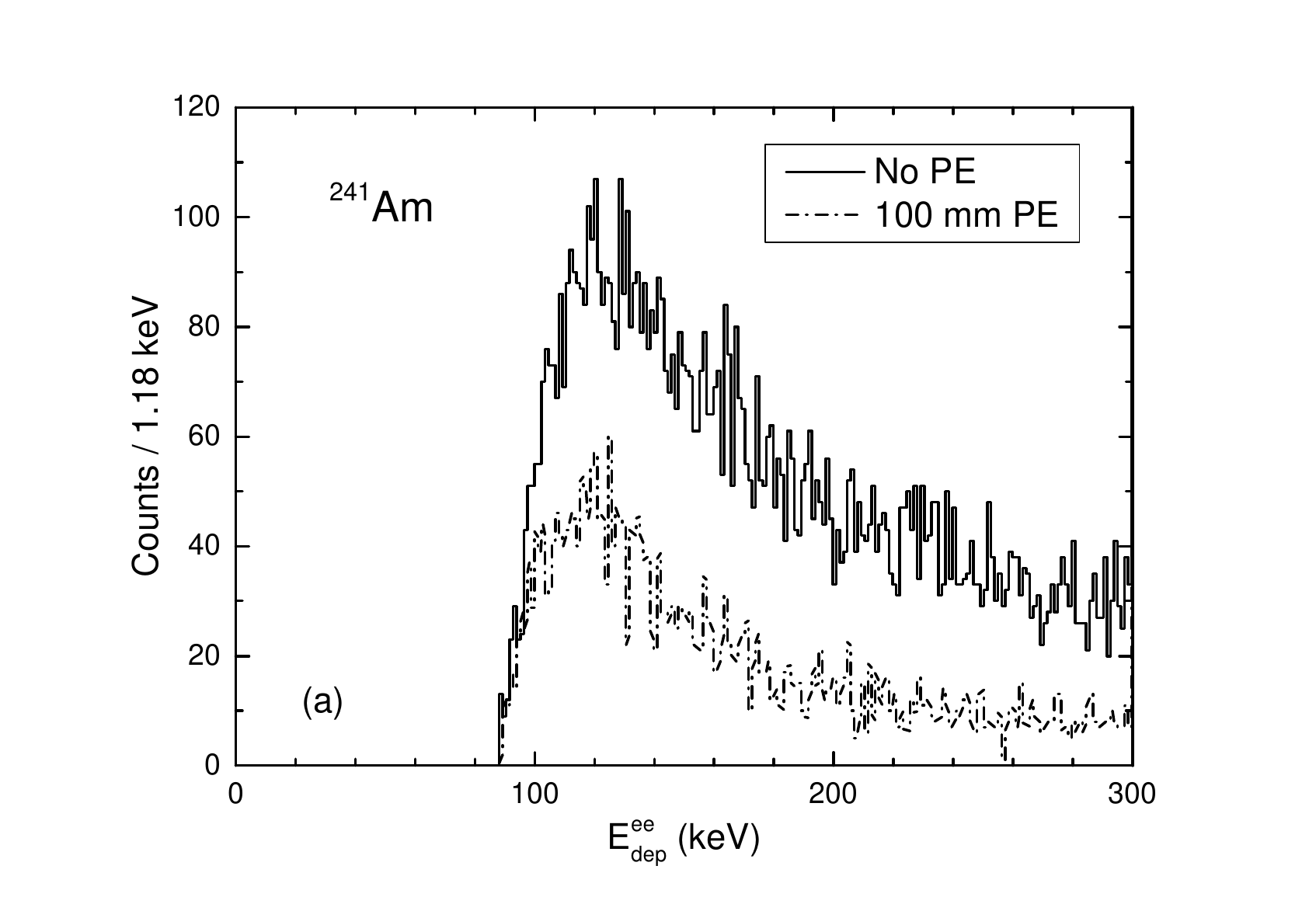}
        \phantomsubcaption
	      \label{fig7a}
    \end{subfigure}%
    \begin{subfigure}[t]{0.5\textwidth}
        \centering
        \includegraphics[height=55mm]{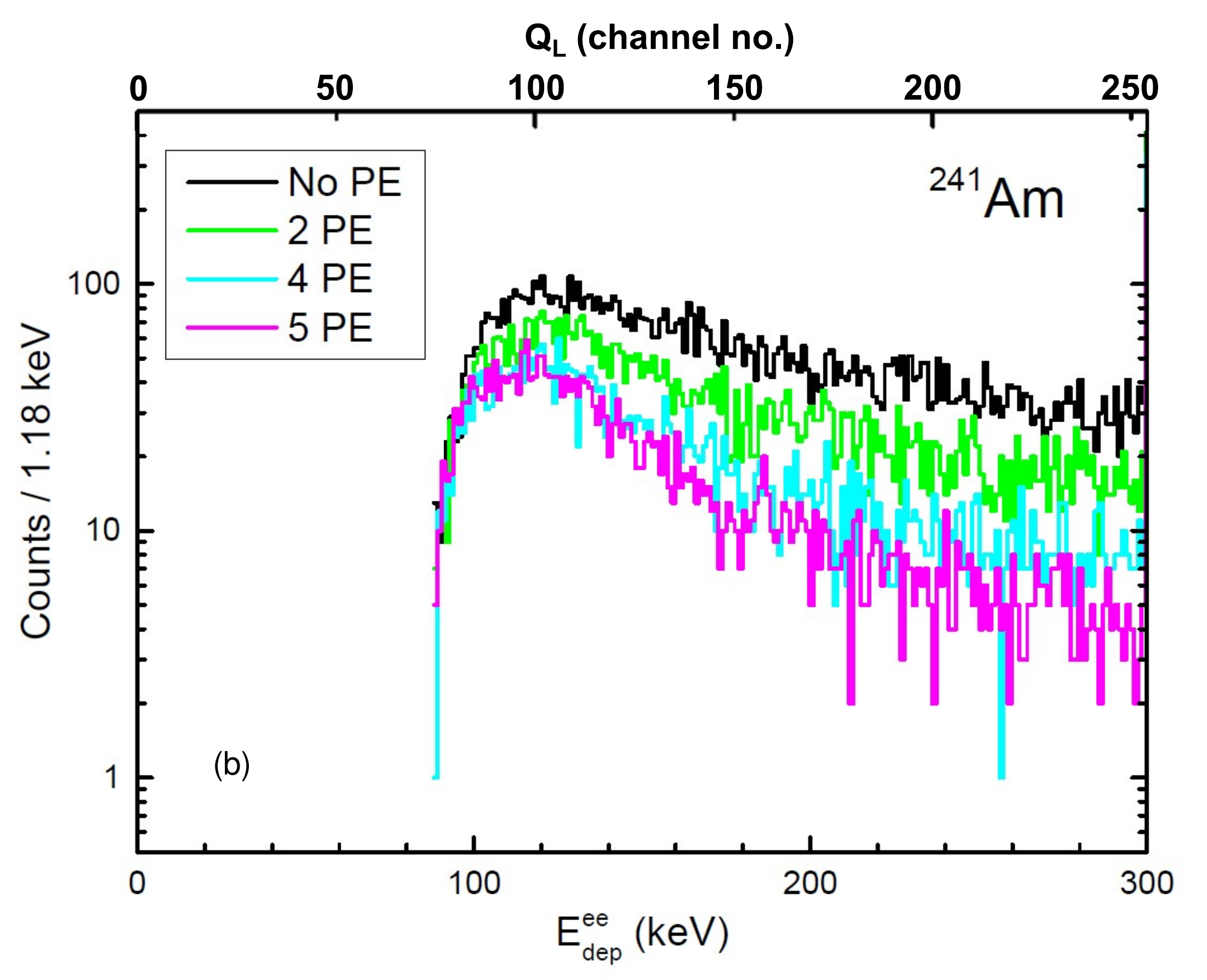}
        \phantomsubcaption
        \label{fig7b}
    \end{subfigure}	
    \vspace{-1.5\baselineskip}
	\caption{(a) Projected spectra on the $Q_L$ axis after setting a banana gate on the neutron band with and without the HDPE layers between the source and the detector. Attenuation of neutrons due to HDPE is evident; (b) comparison of the projected neutron spectra recorded after the neutrons travel through different numbers of the HDPE layers. See Sec.~\ref{HDPEstudies} for the energy calibration applied here.} 
	\label{fig7}
\end{figure}

In order to confirm that the so-called neutron band is due to the neutrons being detected, attenuation of the band population by the HDPE was studied. For this purpose, 4 layers of HDPE, each being 25 mm thick with a total thickness of 100 mm, were placed between the $^{241}$Am source and the PHe detector to attenuate the neutrons. The scatter plot ($P$ vs. $Q_L$), obtained over the same exposure time, is shown in Fig.~\ref{fig6b}.
It clearly reveals attenuation of the neutron band. The plots, obtained by gating on the neutron band and projection taken along the $Q_L$ axis (see Fig.~\ref{fig7}), provide a comparison of the neutron attenuation by the HDPE layers. Low energy neutrons are expected to be attenuated more than those of higher energies, however from the plot, the attenuation factor appears to be fairly independent of energy. This is due to the moderation of the energetic neutrons that enhances counts at low energy and suppresses at higher energy. 

\begin{figure}[htbp!]
	\begin{subfigure}[t]{0.5\textwidth}
        \centering
	    \includegraphics[height=55mm]{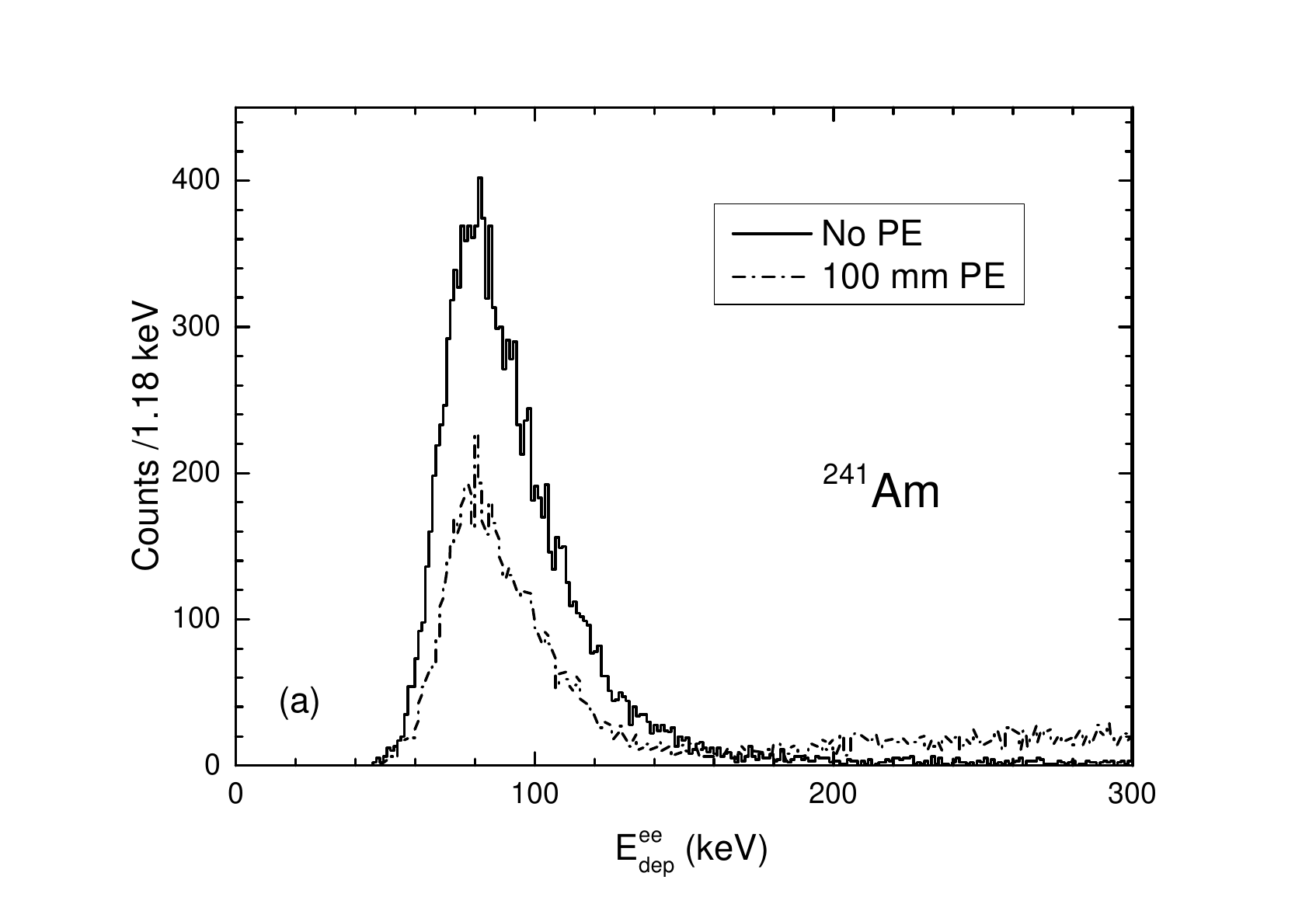}
	    \phantomsubcaption
        \label{fig8a}
    \end{subfigure}%
    \begin{subfigure}[t]{0.5\textwidth}
        \centering
	    \includegraphics[height=55mm]{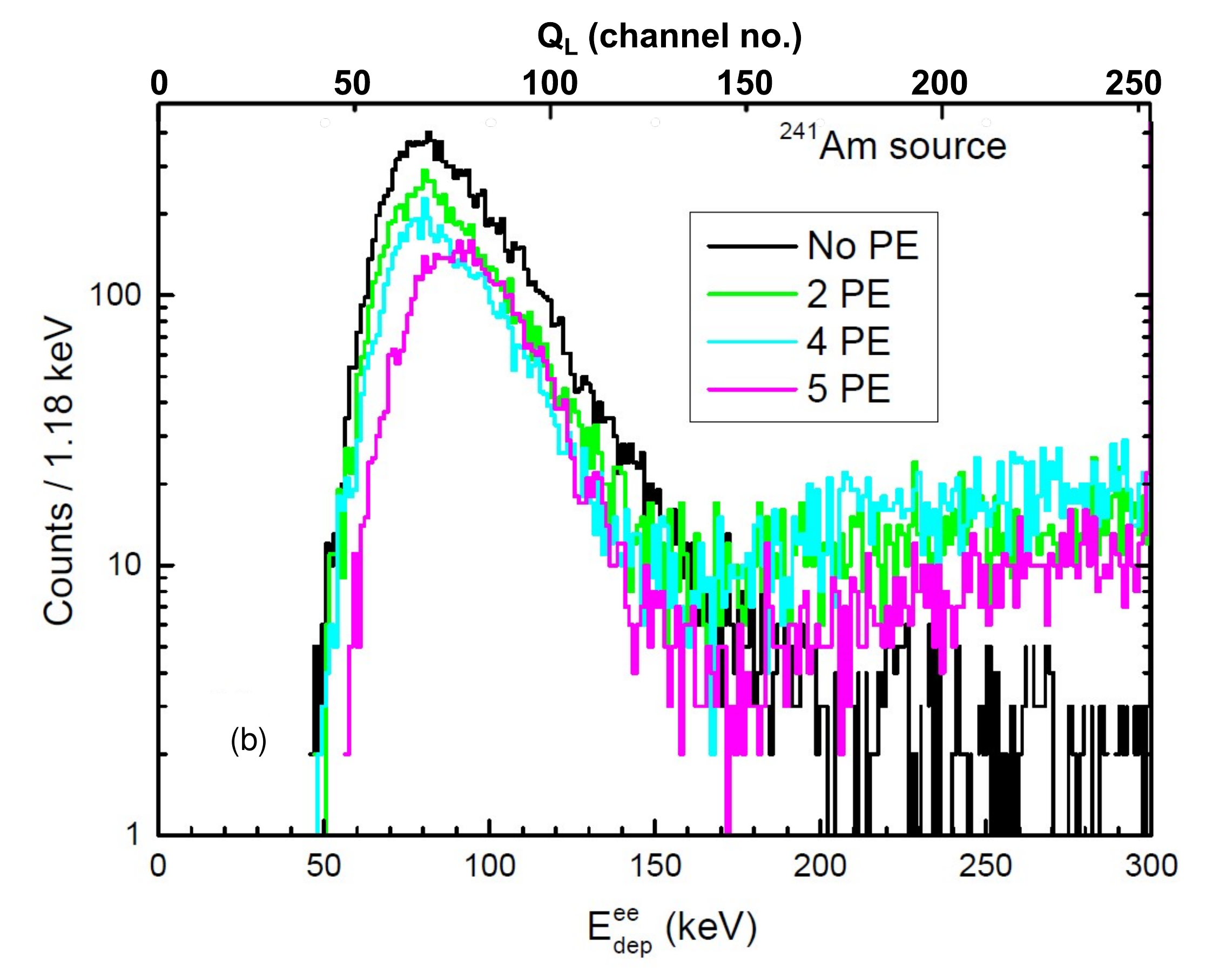}
        \phantomsubcaption
        \label{fig8b}
    \end{subfigure}
    \vspace{-1.5\baselineskip}
	\caption{Same as Fig.~\ref{fig7} except that the banana gate is set on the $\gamma$-band. (a) projected spectra with and without the HDPE (PE) layers between the source and the detector. See Sec.~\ref{HDPEstudies} for the calibration procedure applied in the plot.(b) Comparison of the projected $\gamma$-spectra recorded after the neutrons travel through different numbers of the HDPE layers. The $x$-axis scale on the bottom displays the electron equivalent energy loss $E_{dep}^{ee}$ of neutron. See Sec.~\ref{HDPEstudies} for its definition.} 
	\label{fig8}
\end{figure}

Similar projection spectrum for the $\gamma$-band is shown in the Fig.~\ref{fig8}, where attenuation of $\gamma$-rays is observed. The $\gamma$-ray spectrum at the lower channels along the $x$-axis get significantly attenuated after passing though the HDPE layers due to the presence of scatterers. However, the higher channels of the spectrum corresponding to larger deposited energy ($E_{dep}$) values get populated as well. For measurements with different number of HDPE layers in between, we observe that after traversing two HDPE layers, the spectrum at the higher channels (i.e region of interest or RoI $> 150$) show a rising trend and a bump around 240 channel number till it reaches saturation at the highest channel. There is not much visible relative change for four HDPE layers. However, the spectral population over the RoI diminish after passing through five HDPE layers. This may be interpreted as  due to the production of 2.225 MeV $\gamma$-rays from neutron moderation within the HDPE, followed by absorption in HDPE through the $n + p = d + \gamma$ process. A simulation of the $\gamma$-ray interaction with the detector is done to understand the origin, the nature of the $\gamma$-ray spectra and the underlying systematics of the process.

$\gamma$-rays falling on the PHe detector cause electrons inside the gas to recoil and thus, result in electron recoil signal. This is demonstrated in our experiment (see Fig.~\ref{fig3}) with $^{137}$Cs standard $\gamma$-ray source as well. The range of $E_{dep}$ values can be seen to merge with the lower range of $E_{dep}$ values for the neutrons as shown in the Fig.~\ref{fig4}, where the spectra for the neutrons and the $\gamma$-rays from $^{252}$Cf are overlaid. Henceforth, we will designate the energy loss by the electrons as $E_{dep}^{ee}$ to distinguish it from $E_{dep}$ due to the neutrons. The study is more complicated due to the presence of stainless steel housing of the detector which contains high $Z$ elements. Since the $\gamma$-rays are energetic enough to undergo pair production, positrons are also generated, which result in the production of 511 keV $\gamma$-rays. Therefore, the ER events, though largely contain effects due to the electrons, there would be effect due to the positrons as well.

\begin{figure}[htbp!]
    \centering    
    \includegraphics[height=60mm]{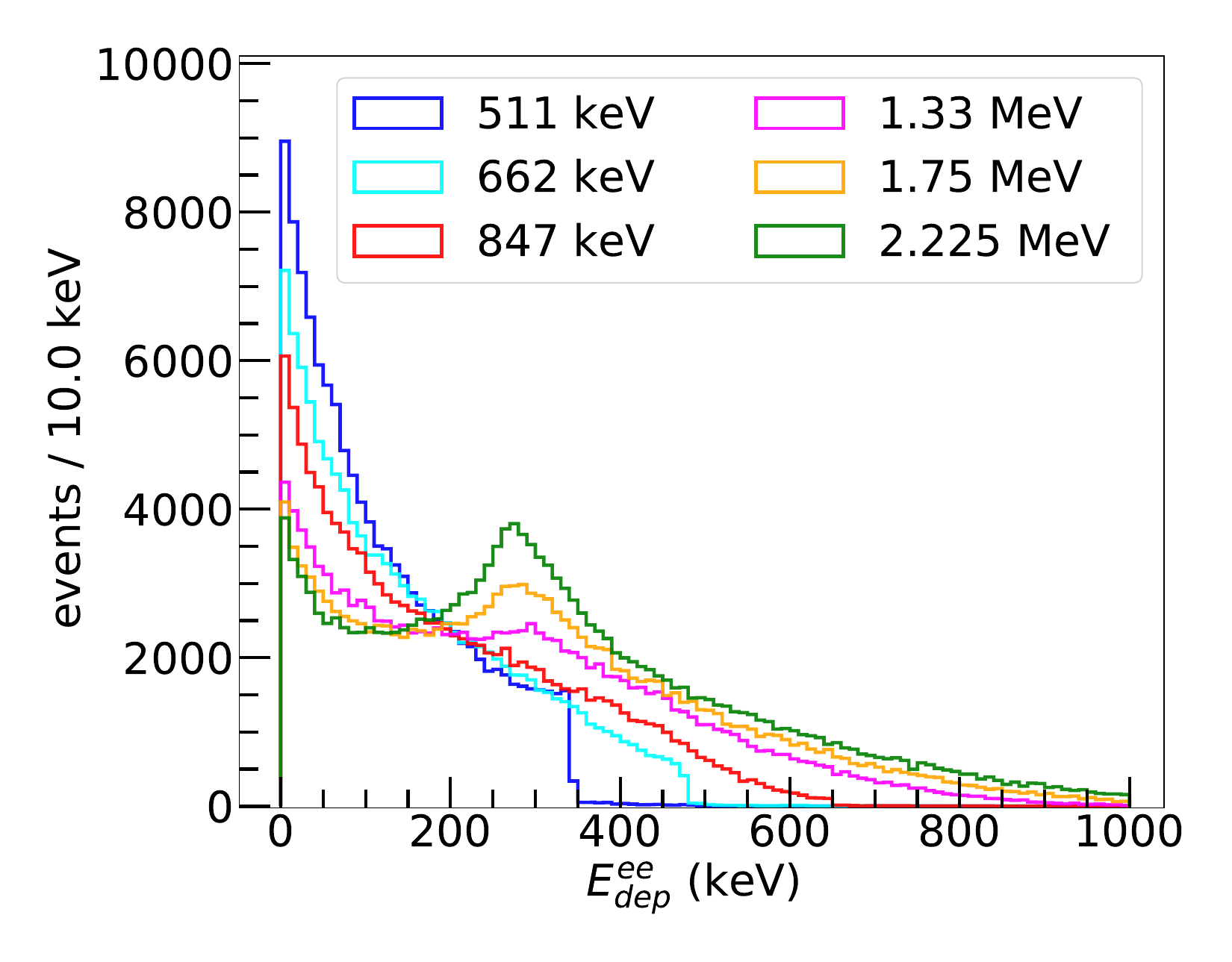}
    \vspace{-1\baselineskip}
    \caption{$E_{dep}^{ee}$ spectrum of mono-energetic $\gamma$-rays}
   \label{fig12a}
\end{figure}

In the simulation procedure, mono-energetic $\gamma$-rays of different energies ($E_\gamma$) are allowed to impinge upon the detector. The $E_{dep}^{ee}$ spectrum of the recoiling electrons is shown in Fig.~\ref{fig12a}. A peak begins to appear at $E_{dep}^{ee} \sim280$~keV as the energy of the incident $\gamma$-ray is increased. The $\gamma$-ray energy threshold for appearance of the peak is found to be around 650~keV. Therefore, the 847~keV $\gamma$-ray, which arises from the $^{56}{\rm Fe}(n,n'\gamma)$ reaction due to the presence of SS housing, would also give rise to the $\sim280$~keV peak, when no HDPE slab is placed. This explains the very few counts observed in the higher channel numbers in the absence of HDPE attenuator in Fig. \ref{fig8b} (black line). A detailed event-by-event tracking reveals that this number corresponds to the most probable $E_{dep}^{ee}$ of the recoiling electrons given the geometry (diameter) of the detector.

\begin{figure}[htbp!]
    \begin{subfigure}{0.5\textwidth}
        \centering    
        \includegraphics[height=60mm]{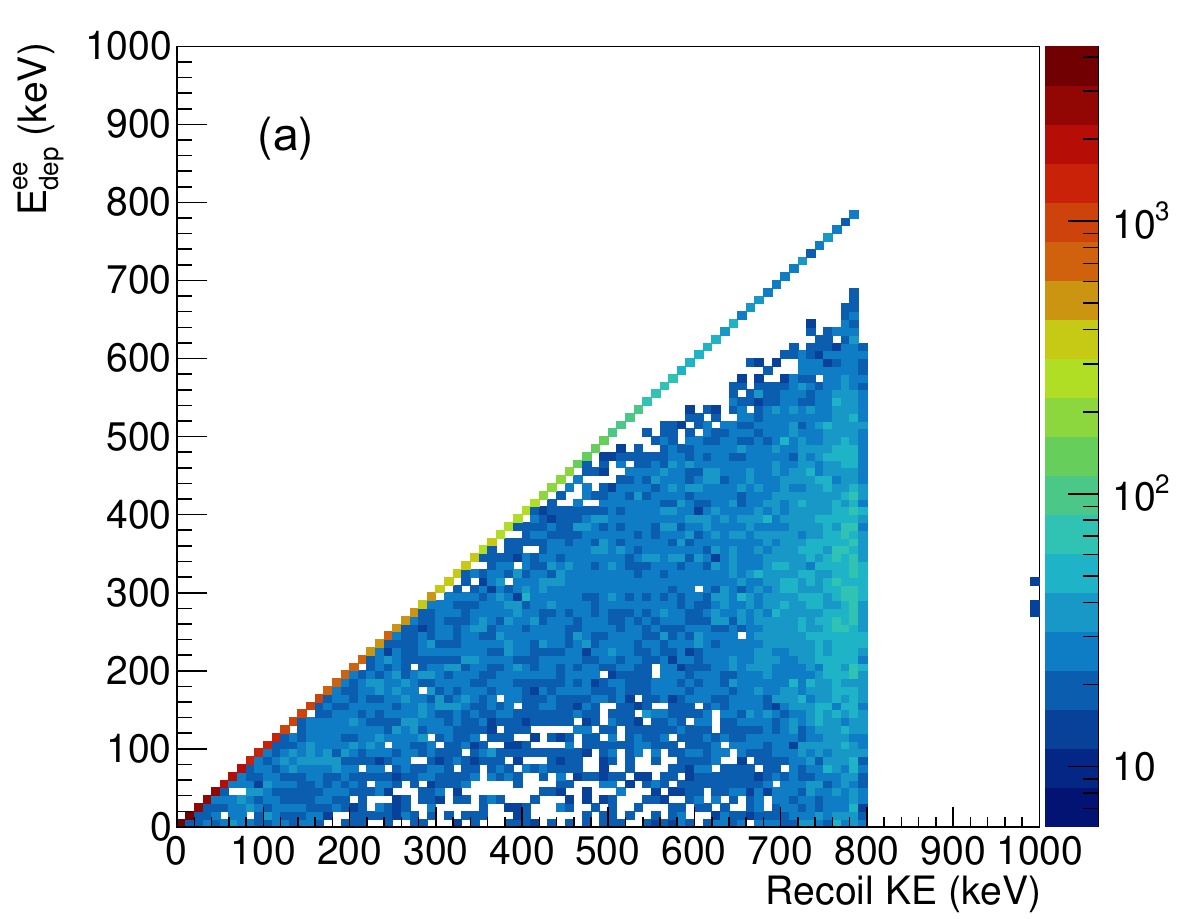}
        \phantomsubcaption
        \label{fig-1p0}
    \end{subfigure}%
    \begin{subfigure}{0.5\textwidth}
        \centering    
        \includegraphics[height=60mm]{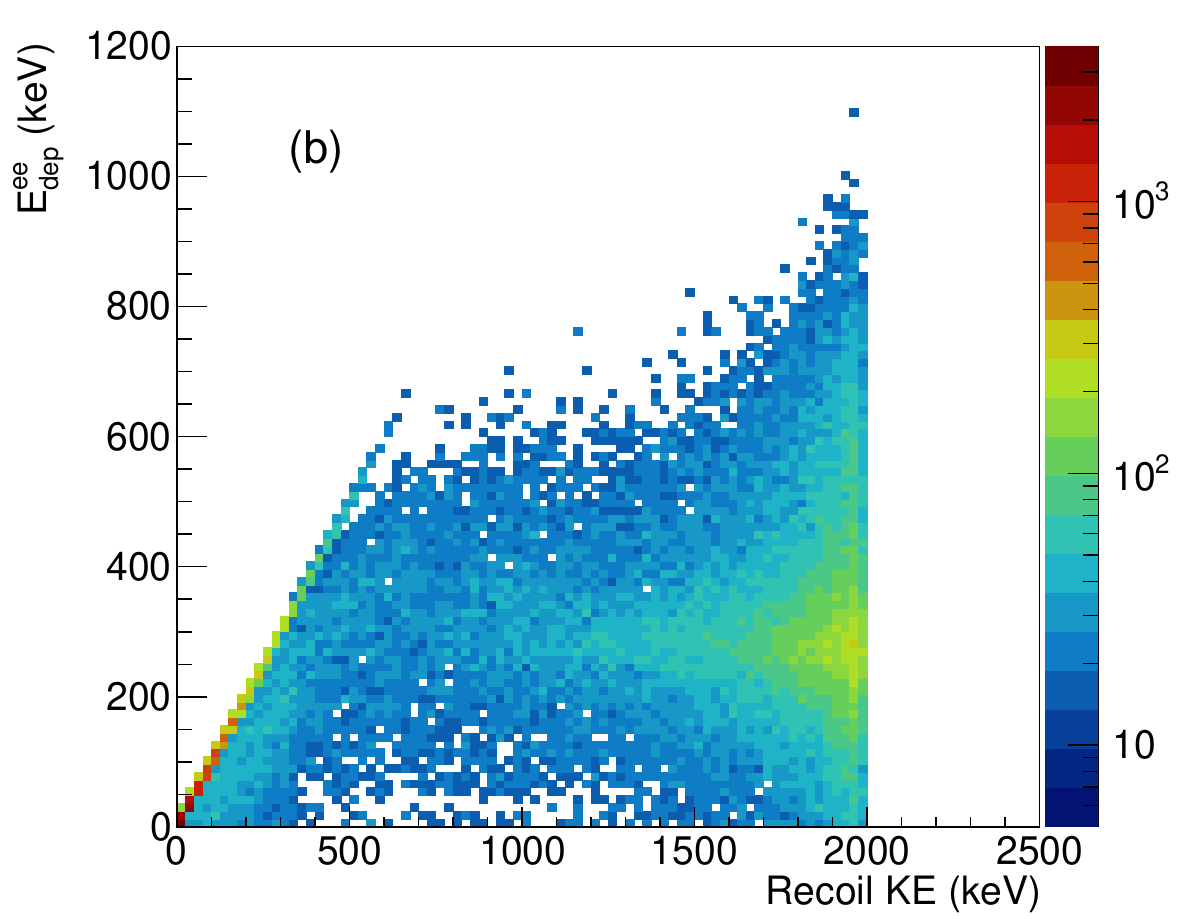}
        \phantomsubcaption
        \label{fig13-2p22}
    \end{subfigure}
    \vspace{-2.5\baselineskip}
    \caption{$E_{dep}^{ee}$ vs. recoil kinetic energy of ER events from mono-energetic $\gamma$ sources for (a) 1.0 MeV and (b) 2.225 MeV}
    \label{fig13}
\end{figure}

\begin{figure}[htbp!]
    \centering    
    \includegraphics[height=60mm]{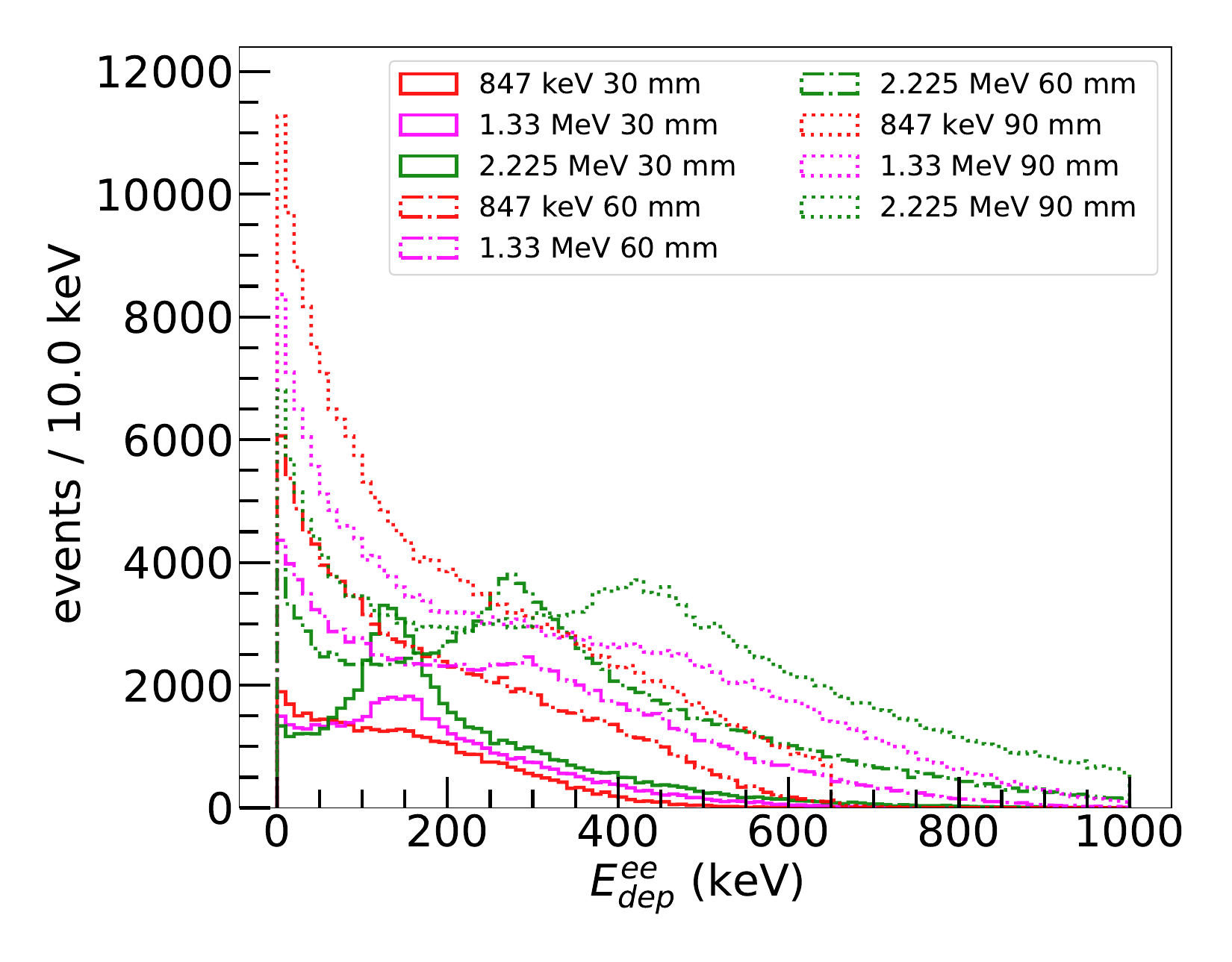}
    \vspace{-1\baselineskip}
    \caption{$E_{dep}^{ee}$ spectra with 30 mm (solid lines), 60 mm (dot-dashed lines) and 90 mm (dotted lines) diameter of the active volume of the detector.}
    \label{fig14}    
\end{figure}

The peak-like feature in the Fig.~\ref{fig12a} arises due to the fact that the recoiling electrons do not lose their entire energy within the detector. This can be seen more explicitly in Fig.~\ref{fig13}, where the $E_{dep}^{ee}$ is plotted against their recoil kinetic energy (KE) for mono-energetic incident $\gamma$-rays of 1~MeV and 2.225 MeV. The kinetic energy of the recoiling electrons increases with incident energy of the $\gamma$-rays as expected. The sharp right edge in the plots is the Compton edge. It is evident from Fig.~\ref{fig13} that a significant fraction of the electrons with higher recoil energies lose only $\sim280$~keV of their energy within the detector. Therefore, we can conclude that the peak observed at higher channel numbers in the Fig.~\ref{fig7b} corresponds to $\sim280$ keV of $E_{dep}^{ee}$. To further confirm this geometric effect, the simulation was repeated by changing the diameter of the fiducial volume of the detector to 30 mm and 90 mm. Corresponding $E_{dep}^{ee}$ spectra are plotted in the Fig.~\ref{fig14}. A clear shift in the peak can be seen, which is at $\sim130$ keV for 30 mm diameter and $\sim450$ keV for 90 mm.

Based on the experimental and the simulation results on the interaction of the neutrons with HDPE and $\gamma$-ray response of the PHe detector, we can conclude that: 1) the build-up observed in the higher channel number is due to the production of 2.225 MeV $\gamma$-rays from $n + p = d + \gamma$ process 2) the $E_{dep}^{ee} \sim280$ keV peak arises due to the most probable value of energy loss by the recoiling electrons within the confined geometry of the PHe detector. This is also seen as a rising bump in the RoI at the higher channel numbers ($\sim240$) in the experiment (see Fig.~\ref{fig8b}), apart from the spectral feature at the lower channel numbers ($\leq \,150$) which has been demonstrated as arising from accompanying low energy $\gamma$-rays. Based on the spectral profile of the $E_{dep}^{ee}$ spectra, we fitted the peak with logistic function after subtracting an exponential background for the $E_{dep}^{ee}$ spectra shown in the Fig.~\ref{fig12a}, covering 662 keV to 2.225 MeV $\gamma$-rays. The average $E_{dep}^{ee}$ value for the peak was found to be: $282 \pm 10$~keV. Considering this and the fact that the $x$-axis of the plots of the Fig.~\ref{fig8b} scale as $E_{dep}^{ee}$, a first order calibration with calibration constant of 1.175 keV/channel on the $x$-axis is applied. 

\subsection{Neutron Energy Calibration based on \texorpdfstring{$^{252}$}{}Cf Spectral Data}
\label{calib}
\begin{figure}[htbp!]
    \begin{subfigure}{0.5\textwidth}	
        \centering
        \includegraphics[height=60mm]{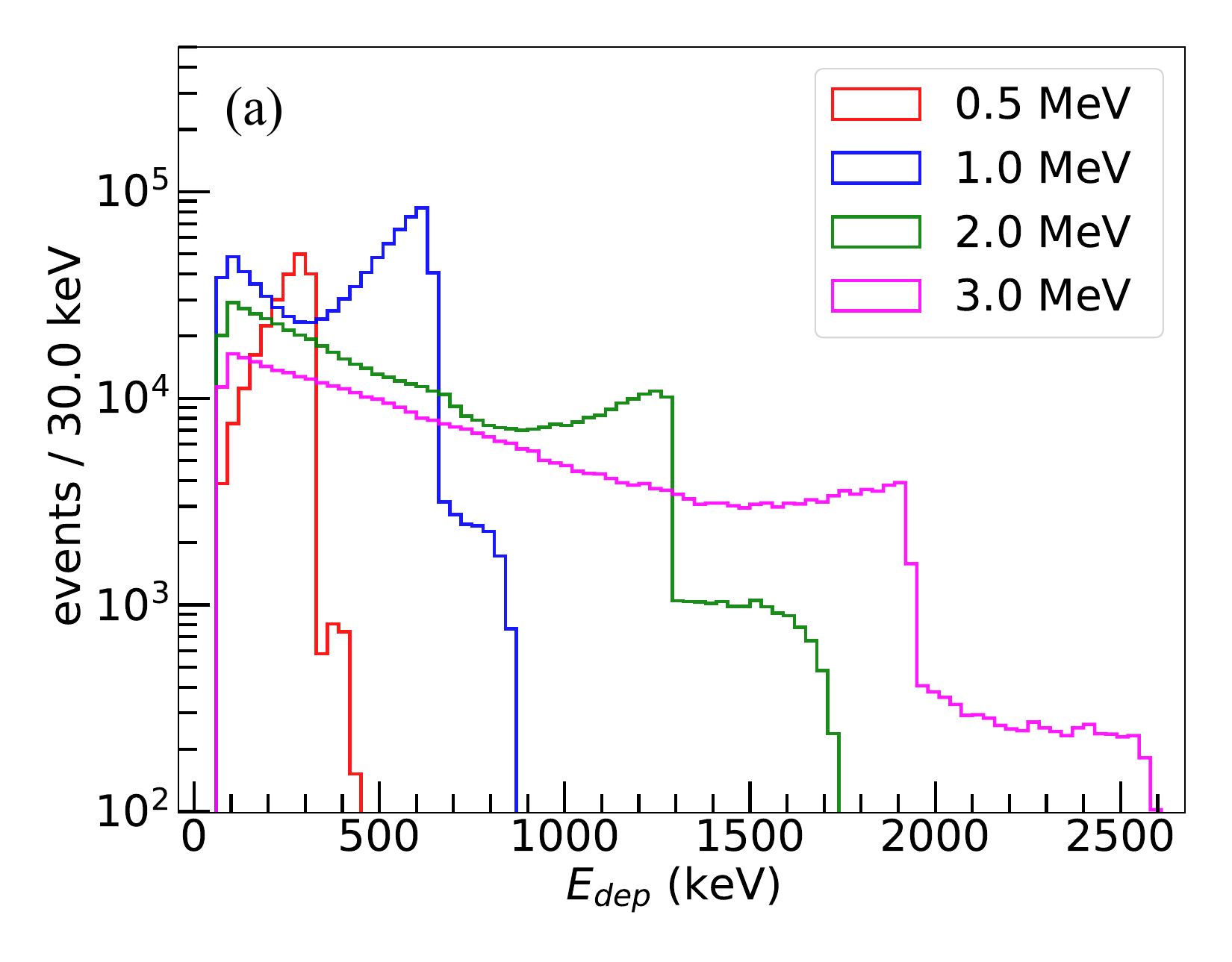}
        \phantomsubcaption
        \label{fig10a}
    \end{subfigure}%
    \begin{subfigure}{0.5\textwidth}	
        \centering
        \includegraphics[height=60mm]{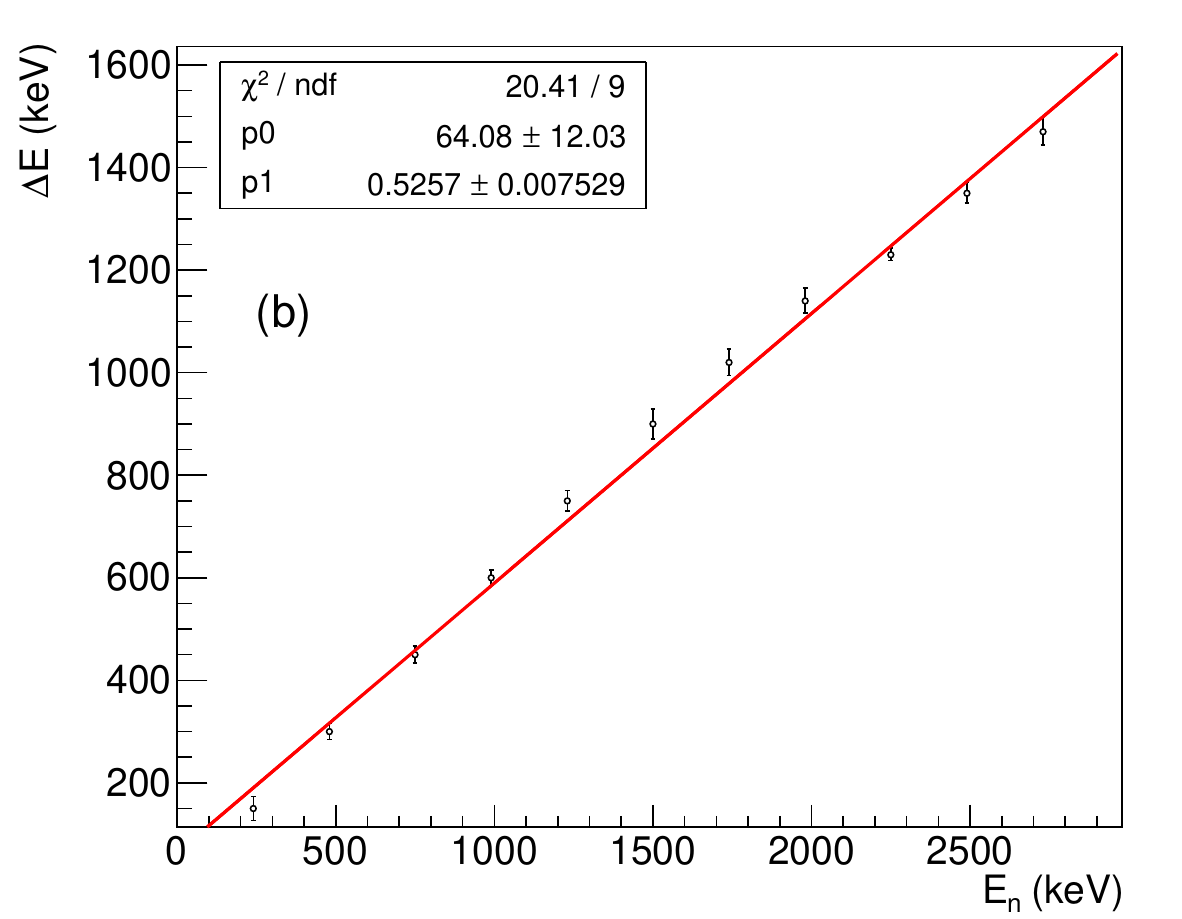}
        \phantomsubcaption
        \label{fig10b}
    \end{subfigure}
    \vspace{-2.5\baselineskip}
    \caption{(a) Simulated $E_{dep}$ spectrum of the mono-energetic neutrons; (b) $\Delta E$ is plotted against energy of the neutrons. See Sec. \ref{calib} for the definition of the $\Delta E$ parameter.}
    \label{fig10}
\end{figure}

As mentioned in Sec.~\ref{Intro}, neutron scatters off a $^4$He nucleus transferring a part of its kinetic energy ($E_n$). The recoiling $^4$He deposits the recoil energy within the fiducial volume of the detector via ionization, followed by scintillation. The recoiling $^4$He deposits its entire energy within the volume, which is designated as $E_{dep}$. The $E_{dep}$ spectra for monoenergetic neutrons at different energies ranging from 0.5 to 3 MeV are estimated as shown in the Fig.~\ref{fig10a}. Since there is spread in the $E_{dep}$ spectra, we have estimated the 90\% value of the total area under the spectral profile. Subsequently, we have determined the upper limit of the $E_{dep}$ parameter for the integral under the spectral profile, which matches with the above-mentioned number. We have designated the corresponding upper limit on $E_{dep}$ as the marker ($\Delta E$) of the corresponding recoil signal registered by the detector. A plot of ($\Delta E$) as function of $E_n$ is shown in the Fig.~\ref{fig10b}. The ($\Delta E$) is found to be fairly linear with the incident energy $E_n$. Therefore, an estimate of the incident energy may also be done from the $E_{dep}$ spectrum.

Based on this observation, an alternative energy calibration of the detector was attempted using the prompt neutrons from a $^{252}$Cf source. It is expected that the energy spectrum of the neutrons originating from the spontaneous fission is Maxwellian. However, after detailed analysis by various groups on the properties of neutron sources, and several meetings under the banner of International Atomic Energy Agency (IAEA), held during 1980 to 1987, the expert group on $^{252}$Cf spontaneous fission-based fast neutron sources, had proposed a corrected Maxwellian spectrum, based on experimental data and theoretical estimates\cite{Mannhart} over various energy segments from 0.2 MeV to 20 MeV. Accordingly, the corrected  Maxwellian spectrum $F(E_n)$ as function of the kinetic energy ($E_n$) of the neutrons for the prompt fission neutrons is given by
\begin{equation}\label{eq1}
F(E_n)=R(E_n)\,\frac{2}{\sqrt{\pi}}\,\frac{\sqrt{E_n}}{T^{3/2}}
\exp [{-\frac{E_n}{T}}],
\end{equation}
where, $R(E_n)$ is the proposed correction factor to the Maxwellian spectrum, and $T$ is the nuclear temperature of $^{252}$Cf before fission, with typical value of 1.42~MeV\cite{Mannhart}. Based on Mannhart's correction factors, a least squares-fitted polynomial regression model was introduced in the Los Alamos ORNL MCNP-DSP code \cite{osti_777654}. The $R(E_n)$ values, based on the above, were used to obtain the corrected Maxwellian form for the $^{252}$Cf prompt neutrons to arrive at the benchmark $^{252}$Cf prompt neutron spectrum. Finally, the simulated spectrum was folded by the intrinsic detector efficiency data available from Ref.~\cite{LIANG20191} by applying cubic spline interpolation. It may be noted that the intrinsic efficiency data, obtained from the time of flight (TOF) measurements, is available in the range of ~0.5 MeV to 6.5 MeV, with backward interpolation extended to 0.35 MeV.

\begin{figure}[ht]
	\centering
	\includegraphics[height=65mm]{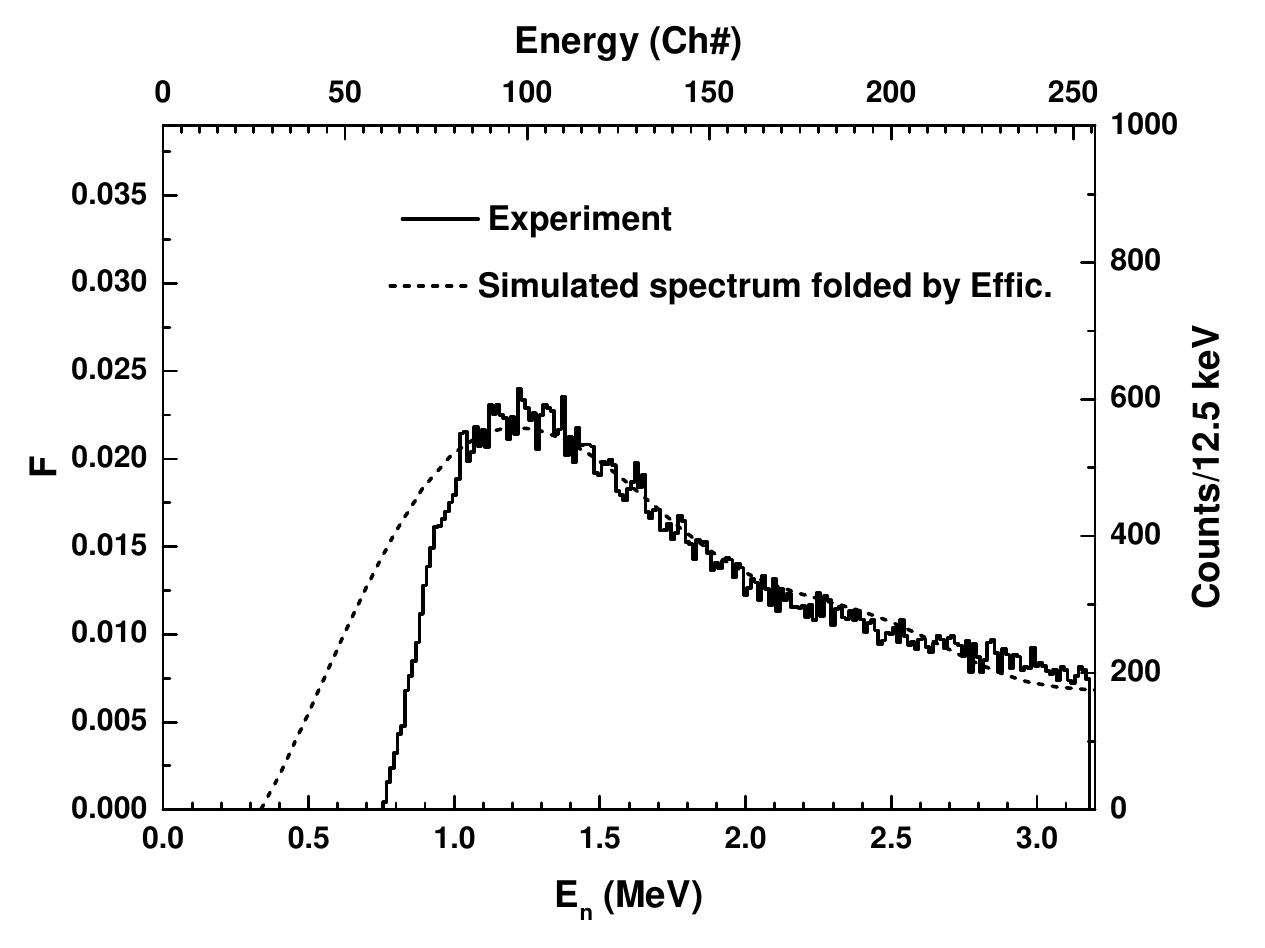}
	\caption{Overlaid experimental and theoretically estimated energy spectra of neutrons from the $^{252}$Cf spontaneous fission source. The top axis label displays the QDC channel number and the bottom axis labels are the energy of the neutrons.}
	\label{fig5}
\end{figure}

The measured and the simulated efficiency values given in Ref.~\cite{LIANG20191} considerably differ over various energy ranges. Therefore, the experimentally obtained spectrum was compared with the theoretical one after folding the latter with: 1) the experimentally obtained efficiencies $\epsilon_{\rm ex}\,({E_n})$, 2) simulated efficiencies $\epsilon_{\rm th}\,({E_n})$, and 3) average of the two efficiencies as mentioned above, estimated at each energy. Our comparisons reveal the best match for the $\epsilon_{\rm ex}\,({E_n})$, which is shown in the Fig.~\ref{fig5}. It is evident that the lower energy part of the spectrum reveals a cut-off around 0.75 MeV, which is possibly contributed by the threshold discriminator setting and also the merging threshold of the neutron and the $\gamma$-bands in our experiment (refer to Fig.~\ref{fig2}). 

A comparison of the experimentally obtained $^{252}$Cf prompt neutron spectrum and the one theoretically obtained as above, are shown in the Fig.~\ref{fig5}. They are found to be in good agreement over the energy range of ~0.7 MeV to ~3.2 MeV. It is important to mention two major points here. a) The upper limit of the available energy range is due to the gain settings of the built-in amplifiers associated with the $^4$He detector. b) Our comparison, shown in the Fig.~\ref{fig5}, manifests that the measured spectrum is constrained by a lower energy cut-off around 0.75 MeV, whereas the estimated spectrum, folded by the efficiency data, has a lower energy cut-off of around 0.35 MeV which arises entirely from the available intrinsic efficiency data in Ref.~\cite{LIANG20191}.

There are uncertainties in this method of calibration, which is primarily caused by the sparse data on the intrinsic efficiency. The measured and the simulated efficiency results considerably differ over various energy ranges\cite{LIANG20191}. Besides, interpolation procedure followed leaves scope for introducing additional uncertainties. Because of these uncertainties, we have adopted a linear calibration without a cut-off. However, a least square fit of the calibration  data up to the quadratic term (ie. $E_n(x) = a_0+a_1 x + a_2 x^2$, where $x$ is the channel number), results in the following best fit values: 
$a_0=0.00167 \pm 0.00212\,{\rm MeV}, a_1=0.0125 \pm 3.93\times 10^{-5}\, {\rm MeV/ch}, a_2=1.87\times 10^{-8} \pm 1.48\times 10^{-7}\,{\rm MeV/ch^2}$. Relatively smaller values of $a_0$ and $a_2$ justify our choice of linear calibration. 

It may be noted that the detector is also capable of detecting the $\gamma$-rays due to its electron recoil response, though the intrinsic detection efficiency may be very small compared to that of the neutrons. Though considerable details about the shape of the energy spectral distribution of $\gamma$-rays from spontaneous fission of $^{252}$Cf is available\cite{Verbke}, similar energy calibration procedure could not be attempted for the $\gamma$-ray spectra, because of the overlap of the neutron and the $\gamma$-bands (see Fig.~\ref{fig2}) at lower energies.

\section{Conclusion}
\label{conclusion}
PHe detector, capable of detecting and discriminating between the neutrons and the $\gamma$-rays in a mixed radiation field, has been evaluated in detail through source-based experiment and supplemented by G4-based simulation.

The experimental studies based on $^{252}$Cf fast neutron source, $^{137}$Cs $\gamma$-reference source and $^{241}$Am source have successfully demonstrated that a fast and a slow gate-based integration of the detector output pulses by the QDC is effective in achieving discrimination. Attenuation of the neutron band by the HDPE layers placed between the source and the detector also confirms the discrimination method followed. The charge contents of the pulses over the long gate effectively scale as the energy deposited $E_{dep}$ by the recoiling $^4$He in case of the neutrons, or the recoiling electrons in case of the $\gamma$-rays. 

Parallel measurement of the $\gamma$-ray spectral profiles for the radioactive sources mentioned above were also done. The $\gamma$-ray spectra, arising due to the scintillation caused by the corresponding electron recoil events, reveal peak-like structures at the larger $E_{dep}$ values, which is grossly correlated with the placement of HDPE absorbers between the source and the detector.

Interaction of the neutrons and the $\gamma$-rays with the HDPE layers was investigated in detail through G4-based simulation. Production of the neutrons from the $^{241}$Am source was demonstrated to be largely due to the ($\alpha,n$) reactions on the constituent elements of the glass substrate. Corresponding neutron spectrum was used for the event generation. Interaction of the $\gamma$-rays emitted from the $^{241}$Am source and also produced through neutron absorption by the HDPE layers with the pressurized Helium active medium, were studied to find the origin of the peak-like structures in the high-end tail of the measured spectra. It is found that a significant number of relatively higher energy $\gamma$-rays ($E_\gamma \gtrsim 650\,{\rm keV}$) produce recoiling electrons, which dump around 280 keV of energy, resulting in the peak-like structure. This observation was used to derive a first-order calibration in terms of $E_{dep}^{ee}$, which was then applied to the measured neutron spectra. Role of the detector geometry was investigated as part of the systematic studies in support of the above.

\section{Acknowledgements}
We would like to acknowledge technical and operational support from Chandranath Marick of SINP, India. One of us (SS) would like to acknowledge financial support from the Department of Atomic Energy Raja Ramanna Fellowship (DAE-RRF) scheme to carry out this work.

\bibliographystyle{JHEP} 
\bibliography{references}

\end{document}